\documentclass[prx,twocolumn,10pt,aps,floatfix]{revtex4-2}

\usepackage[dvips]{graphicx}
\usepackage{epsfig}

\usepackage[margin=2cm,head=0.5cm]{geometry}
\usepackage[latin1]{inputenc}
\usepackage{bm}
\usepackage{amsmath}
\usepackage{amssymb}
\usepackage{latexsym}
\usepackage{amsfonts}
\usepackage{epsfig}
\usepackage{color}
\usepackage[linktocpage, colorlinks=true ,linkcolor=blue, citecolor=blue]{hyperref}
\usepackage[all]{hypcap}
\usepackage{orcidlink}

\newcommand{\expval}[1]{\left< #1 \right>}

\newcommand{\ket}[1]{\left|#1\right>}
\newcommand{\bra}[1]{\left<#1\right|}
\newcommand{\braket}[2]{\left<#1|#2\right>}

\newcommand{\nn}{\nonumber\\}

\newcommand{\f}[1]{\mbox{\boldmath$#1$}}

\newcommand{\bea}{\begin{eqnarray}}
\newcommand{\eea}{\end{eqnarray}}
\newcommand{\beann}{\begin{eqnarray*}}
\newcommand{\eeann}{\end{eqnarray*}}
\newcommand{\ord}{{\cal O}}
\newcommand{\trace}[1]{{\rm Tr}\left\{ #1 \right\}}
\newcommand{\traceB}[1]{{\rm Tr_B}\left\{ #1 \right\}}
\newcommand{\traceS}[1]{{\rm Tr_S}\left\{ #1 \right\}}
\newcommand{\abs}[1]{{\left| #1 \right|}}

\newcommand{\ii}{\mathrm{i}}  

\newcommand{\rev}[1]{#1} 
\definecolor{blau}{RGB}{0,70,180}
\definecolor{gruen}{RGB}{0,180,0}

\begin{document}
  
\title{Floquet analysis of a superradiant many-qutrit refrigerator}

\author{Dmytro Kolisnyk\orcidlink{0000-0002-8612-8202}$^1$}
\author{Friedemann Quei{\ss}er\orcidlink{0000-0001-7378-0851}$^2$}
\author{Gernot Schaller\orcidlink{0000-0003-0062-9944}$^2$}
\email{g.schaller@hzdr.de}
\author{Ralf Sch\"utzhold$^{2,3}$}
\affiliation{$^1$Institute of Science and Technology Austria, Am Campus 1, 3400 Klosterneuburg, Austria}
\affiliation{$^2$Helmholtz-Zentrum Dresden-Rossendorf, Bautzner Landstra\ss e 400, 01328 Dresden, Germany}
\affiliation{$^3$Institut f\"ur Theoretische Physik, Technische Universit\"at Dresden, 01062 Dresden, Germany}

\date{\today}

\begin{abstract}
We investigate superradiant enhancements in the refrigeration performance of a set of $N$ three-level systems that are collectively coupled to a hot and a cold thermal reservoir and are additionally subject to collective periodic (circular) driving. 
Assuming the system-reservoir coupling to be weak, we explore the regime of stronger periodic driving strengths by comparing
collective weak driving, Floquet-Lindblad, and Floquet-Redfield master equations.
We identify regimes where the power injected by the periodic driving is used to pump heat from the cold to the hot reservoir and derive analytic sufficient conditions for them based on a cycle analysis of the Floquet-Lindblad master equation.
In those regimes, we also argue for which parameters collective enhancements like a quadratic scaling of the cooling current with $N$ can be expected and support our arguments by numerical simulations.
\end{abstract}

\maketitle
\section{Introduction}

The interaction with reservoirs normally tends to destroy the fragile properties of quantum systems, such that in standard quantum computation applications~\cite{nielsen2000}, decoherence and dissipation~\cite{breuer2002,schlosshauer2007,rivas2012} is often seen as an enemy.
The reservoir-induced decay to the steady state may however exhibit interesting (transient) quantum features on its own, 
as is known from Dicke superradiance~\cite{dicke1954a,gross1982a,benedict1996a}: The decay characteristics of multiple open quantum systems that are identically coupled to a common reservoir may be substantially different from isolated ones.
Studies of open systems have also revealed that they may for multiple reservoirs also function as heat engines~\cite{alicki1979a}, which has fostered the emergence of an entire research field quantum thermodynamics~\cite{binder2019}.
To study open systems as heat engines, one may consider finite-stroke engines that mimic classical thermodynamic cycles by alternatingly coupling to different thermal reservoirs like in a quantum Otto cycle~\cite{kosloff2017a}.
As it may be challenging to completely isolate such quantum working fluids from selected environments in a controlled fashion, an alternative approach uses setups where the working fluid is constantly coupled to multiple reservoirs without additional driving~\cite{kosloff2014a}.
Then, for setups with at least three reservoirs, it may e.g. be possible to cool the coldest of them~\cite{levy2012b} by absorbing heat from the hottest.
However, experimentally, the realization of three reservoirs at vastly different temperatures coupled to the very same quantum working fluid without direct heat flows between the reservoirs may also appear challenging.
Therefore, as a compromise one may imagine quantum heat engines that are subject to both time-dependent periodic driving and are simultaneously coupled to two reservoirs~\cite{szczygielski2013a}.

The general question of collective enhancements~\cite{mukherjee2021a,yadin2023a,filho2023a,rolandi2023a,mamede2023a} of engine performance has been addressed in these engine types, including examples for finite-stroke engines~\cite{hardal2015a,uzdin2016a,cakmak2016a,kloc2019a,watanabe2020a,tajima2021a,kamimura2022a,jaseem2023a,jaseem2023b,eglinton2023a} and continuously operating engines~\cite{manzano2019a,kloc2021a,kolisnyk2023a}.
Also periodically driven collective engines have been treated with weak-amplitude master equations~\cite{macovei2022a,da_silva_souza2022a}.
However, for the last case we remark that the interesting regime of stronger driving amplitudes requires a proper Floquet treatment that has to our knowledge only been explored for self-commuting forms of driving in collective heat engines~\cite{niedenzu2015a,niedenzu2018a}.
The present paper attempts to close this descriptive gap.

We proceed by introducing our model system below in Sec.~\ref{SEC:model}, after which we briefly expose our methods in Sec.~\ref{SEC:methods}.
We then directly discuss our results for a single heat engine and collective enhancements in Sec.~\ref{SEC:results} before concluding.
\rev{We use a standard theoretical apparatus to analyze a collective system of qutrits, subject to periodic driving and dissipative coupling to thermal reservoirs.
However, as what is considered ''standard'' is different in scientific sub-communities, we provide more technical details e.g. on the microscopic derivation of the used master equations specific for our model, the inclusion of energy counting statistics and the collective spin basis in several appendices.}

\section{Model}\label{SEC:model}

\begin{figure}
    \centering
    \includegraphics[width=0.4\textwidth]{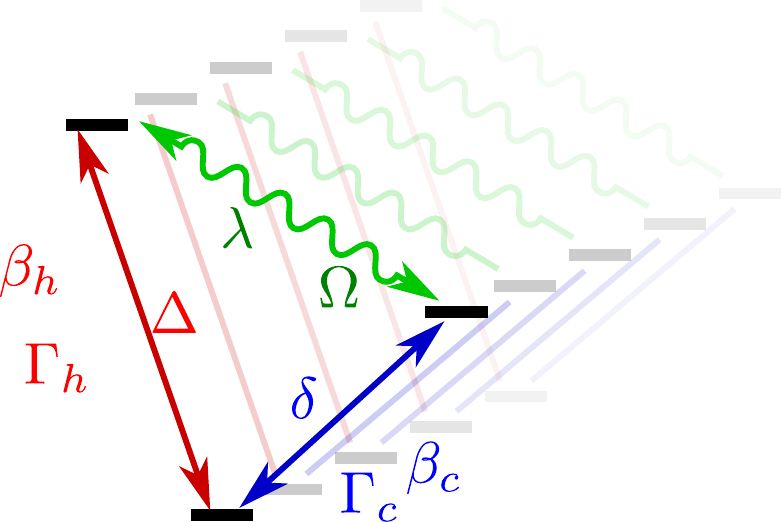}
    \caption{Sketch of the model: A qutrit coupled to hot and cold reservoirs via selective transitions (red and blue arrows, respectively) can act as a refrigerator for the cold reservoir when periodic driving (green wavy line) is additionally applied. For multiple qutrits coupled and driven collectively, superradiant enhancements may occur.}
    \label{FIG:qutrits_driven}
\end{figure}

Our quantum working fluid (compare also Fig.~\ref{FIG:qutrits_driven}) can be described by $N$ identical qutrits, described by states $\ket{0/1/2}$ each, where the energy above the ground state is $\delta$ and the energy of the second excited state is $\Delta>\delta$
\begin{align}
    H_S^0 &= \Delta \sum_{i=1}^N (\ket{2}\bra{2})_i + \delta \sum_{i=1}^N (\ket{1}\bra{1})_i\,,
\end{align}
and where we have gauged the ground state energy to zero.
The qutrits are collectively coupled to two thermal reservoirs held at inverse temperatures $\beta_\nu$:
A hot reservoir ($\nu=h$), which exclusively induces the large transition, and a cold reservoir ($\nu=c$), which exclusively induces the small transition, by using the coupling Hamiltonian
\begin{align}
    H_I^\nu &= J^\nu \otimes B^\nu\,,\nn
    J^c &= \sum_i (\ket{1}\bra{0})_i + {\rm h.c.} \equiv J_c^+ + J_c^-\,,\nn
    J^h &= \sum_i (\ket{2}\bra{0})_i + {\rm h.c.} \equiv J_h^+ + J_h^-\,,
\end{align}
where $B^\nu$ are generic reservoir operators (that will later-on determine the spectral coupling density or spectral function of the reservoir).
As the terminology suggests, we consider cases where $\beta_c \ge \beta_h$, but our methods do of course also apply to the opposite case.
The remaining transition is circularly driven
\begin{align}
    V(t) &= \lambda e^{+\ii\Omega t} \sum_i (\ket{1}\bra{2})_i + {\rm h.c.}\nn
        &\equiv \lambda e^{+\ii\Omega t} J_w^- + \lambda^* e^{-\ii \Omega t} J_w^+\,,
\end{align}
with frequency $\Omega$ and driving strength $\lambda$.
Such drivings could be realized by coupling the transition to a (spatially homogeneous) circularly polarized electromagnetic field~\cite{grifoni1998a,eckardt2015a,bukov2015a,mori2023a}.
We will consider cases where $\Omega<\Delta-\delta$ (red-detuned), $\Omega=\Delta-\delta$ (resonant), and $\Omega>\Delta-\delta$ (blue-detuned) and also explore the role of stronger driving strengths $\lambda$.
Denoting the standard bosonic reservoir Hamiltonians by $H_B^\nu$, the complete Hamiltonian of our model universe is thus given by
\begin{align}
    H(t) &= H_S(t) + \sum_\nu H_I^\nu + \sum_\nu H_B^\nu\,,\nn
    H_S(t) &= H_S^0 + V(t)\,,
\end{align}
which cannot be solved exactly in general, such that we will have to use perturbative methods.  
Thus, in essence, our model represents a collective generalization of a three-level maser~\cite{scovil1959a,boukobza2006a,jaseem2020a,PhysRevA.96.063806}, where we are however primarily interested in the refrigerator operational mode~\cite{boukobza2007a}.

Our first question are the conditions under which it is in the long run (discarding any initial relaxation behaviour) possible to cool the cold reservoir, i.e., whether by investing work via the driving $V(t)$ one can transport energy out of the cold reservoir.
Let us first consider some simple cases that can be understood for $N=1$:
First, in absence of driving ($\lambda=0$), the cold reservoir can only induce small excitations, but the energy cannot be transferred to the hot reservoir as the transition $\ket{1}\to\ket{2}$ is not possible in this case. 
The same situation arises for finite $\lambda$ but $\Omega\to\infty$, which allows to perform a rotating wave approximation on the Hamiltonian. 
Therefore, to leading order in the coupling, the reservoirs can not exchange energy in these limits and the long-term energy current must vanish, leaving no chance for cooling functionality.
Second, for $\Omega=0$ and finite $\lambda$ we obtain an undriven system, where however the internal system dynamics allows for coherent transitions between the two excited states.
In this case, heat flow is possible in the long-term limit, but the second law of thermodynamics tells us that it will always be directed from hot to cold and thus not provide cooling functionality (corresponding to a negative cooling current in our conventions).
Third, for $\lambda\to\infty$, the eigenstates of $H_S(t)$ approach those of $V(t)$ with eigenstates $\ket{0_v}=\ket{0}$ and $\ket{\pm_v}\propto \ket{1} \pm e^{-\ii\Omega t} \ket{2}$.
Fermi's golden rule then suggests that both reservoirs trigger the transitions between the system eigenstates in the same way
$\abs{\bra{0} J^\nu \ket{\pm}}^2 = 1/2$, such that the energy flows symmetrically into both hot and cold reservoirs.
In this case, we simply expect that the driving will heat up both reservoirs.
These considerations suggest that the identification of a regime of cooling functionality deserves a more detailed analysis.

Once such a cooling functionality is established, our second question will be whether collective effects (analogous to superradiance~\cite{dicke1954a,gross1982a,benedict1996a}) can improve the cooling performance of the refrigerator:
This is already the case when the cooling current of $N$ collectively coupled qutrits is larger than $N$ times the cooling current of a single qutrit device, but we are also interested in the scaling behaviour with $N$.
To keep the problem treatable, we therefore assume that the complete symmetry under qutrit permutations is also respected by the initial state (take e.g. $\ket{0\ldots 0}$), which allows us to consider only the permutationally symmetric sector of the dynamics.

\section{Methods}\label{SEC:methods}

Below, we introduce different levels of description, where in the main text we mainly state their mode of application and in the appendices we provide details on their microscopic derivation, thermodynamic discussion and further implications.

\subsection{Weak-driving Lindblad equation}

The weak-driving Lindblad~\cite{lindblad1976a,gorini1976a} equation can be formally obtained by using the dissipators for $\lambda=0$ and replacing only the Hamiltonian by its driven version~\cite{kalaee2021a,macovei2022a}.
In App.~\ref{APP:weakdriving} we provide a microscopic derivation, from which it becomes apparent that this procedure is valid for small $\lambda$ only.
The explicit form of the weak-driving master equation is
\begin{align}\label{EQ:lindblad_adiabatic}
    \dot\rho &= -\ii [H_S^0 + V(t), \rho] + {\cal L}_c \rho + {\cal L}_h \rho\,,\nn
    {\cal L}_\nu \rho &= \gamma_\nu(+\Omega_\nu) \left[J_\nu^- \rho J_\nu^+ - \frac{1}{2}\left\{J_\nu^+ J_\nu^-, \rho\right\}\right]\nn
    &\qquad+\gamma_\nu(-\Omega_\nu) \left[J_\nu^+ \rho J_\nu^- - \frac{1}{2}\left\{J_\nu^- J_\nu^+, \rho\right\}\right]\,,
\end{align}
where the Fourier transform of the reservoir correlation function~\rev{\eqref{EQ:corrfunc}}
\begin{align}\label{EQ:rescorrfunc}
    \gamma_\nu(\omega) = \Gamma_\nu(\omega) [1+n_\nu(\omega)]
\end{align}
can be decomposed in a spectral coupling density (or spectral function) $\Gamma_\nu(\omega)$ (that is positive for $\omega>0$ and depends on the $B^\nu$ operators) and the Bose distribution $n_\nu(\omega) = \frac{1}{e^{\beta_\nu\omega}-1}$ (that imprints the reservoir temperatures on the dynamics) and is evaluated at the original transition frequencies of the quantum working fluid $\Omega_c = \delta$ and $\Omega_h=\Delta$.
As we chose the analytic continuation $\Gamma_\nu(-\omega)=-\Gamma_\nu(+\omega)$, it follows that $\gamma_\nu(-\Omega_\nu) = \Gamma_\nu(\Omega_\nu) n_\nu(\Omega_\nu) \ge 0$, making the Lindblad form explicit and also demonstrating that the isolated dissipators ${\cal L}_\nu$ alone would have the system Gibbs state of the undriven system at their respective reservoir temperature as a steady state.

For systems subject to time-dependent driving, defining the currents properly is in general non-trivial~\cite{gelbwaser_klimovsky2015c,brandner2016a}.
For example, by looking at the energy balance of the bare system $H_S^0$ only, one can obtain energy currents entering the system from the reservoirs as
$I_E^\nu(t) = \trace{H_S^0 ({\cal L}_\nu \rho(t))}$
and the power $P(t) = -\ii \trace{[H_S^0, V(t)] \rho(t)}$, which leads to a consistent thermodynamic description~\cite{kalaee2021a}.
Additionally, it complies with the idea that for small $\lambda$, the energy of the bare system Hamiltonian $H_S^0$ is most relevant. 
However, thinking of device performance (refrigeration), what matters is not the energy balance of the (bare) system, but that of the (cold) reservoir. 
Therefore, we propose to use the counting field formalism, which upgrades ${\cal L}_\nu \to {\cal L}_\nu(\chi_\nu)$ with energy counting fields $\chi_\nu$ via the replacements 
\begin{align}\label{EQ:replacement1}
    J_\nu^- \rho J_\nu^+ \to J_\nu^- \rho J_\nu^+ e^{+\ii\Omega_\nu \chi_\nu}\,,\nn
    J_\nu^+ \rho J_\nu^- \to J_\nu^+ \rho J_\nu^- e^{-\ii\Omega_\nu\chi_\nu}\,.
\end{align}
We stress that this replacement actually results from a microscopic derivation involving the statistics of reservoir energy changes, which we sketch in App.~\ref{APP:fcs} for all the master equations we use.
Within the counting field formalism, the energy currents leaving the reservoirs (by our conventions positive when decreasing the reservoir energy) are then obtained as
\begin{align}
    I_E^\nu(t) &= +\ii \trace{{\cal L}_\nu'(0) \rho(t)}\\
    &= \Omega_\nu [\gamma_\nu(-\Omega_\nu) \trace{J_\nu^- J_\nu^+ \rho} - \gamma_\nu(\Omega_\nu) \trace{J_\nu^+ J_\nu^- \rho}]\,.\nonumber
\end{align}
Using $J_c^+ N_\delta J_c^- - \frac{1}{2}\{J_c^+ J_c^-, N_\delta\} = - J_c^+ J_c^-$ and $J_c^- N_\delta J_c^+ - \frac{1}{2} \{J_c^- J_c^+, N_\delta\} = + J_c^- J_c^+$, one can show that for the weak-driving master equation the above energy currents are indeed identical with the previous definition based on the energy balance of $H_S^0$.
Furthermore, we show in the appendix that there exists a frame where the generator of the weak-driving master equation becomes time-independent and that the currents defined this way actually settle to a steady-state value.
The power at steady state can then be obtained via $\bar P = -\bar I_E^c - \bar I_E^h$ by invoking the first law or by using the definition above, and we also provide in App.~\ref{APP:weakdriving} a short discussion of thermodynamic consistency.

\subsection{Floquet-Lindblad equation}

To investigate the region of larger driving strengths $\lambda$, we use an exact Floquet representation of the system time evolution operator (defined by $\dot U_S = -\ii [H_S^0 + V(t)] U_S(t)$).
The Floquet-Lindblad master equation~\cite{mori2023a} can then be derived microscopically (see App.~\ref{APP:floquet}) under the usual Born-Markov assumptions and a secular approximation relying on vastly different Floquet energy differences and driving frequency $\Omega$~\cite{bulnes_cuetara2015a}.
In the system interaction picture (denoted by bold symbols), where $\f{\rho}(t) = U_S^\dagger(t) \rho(t) U_S(t)$, it can be written as 
\begin{align}\label{EQ:floquet_lindblad}
\dot{\f{\rho}} &= \sum_a \gamma_c\big(-\epsilon_a+\frac{\Omega}{2}\big) \left[J_{c,a}^+ \f{\rho} J_{c,a}^- - \frac{1}{2} \left\{J_{c,a}^- J_{c,a}^+, \f{\rho}\right\}\right]\nn
    &\quad+\sum_a \gamma_c\big(+\epsilon_a-\frac{\Omega}{2}\big) \left[J_{c,a}^- \f{\rho} J_{c,a}^+ - \frac{1}{2} \left\{J_{c,a}^+ J_{c,a}^-, \f{\rho}\right\}\right]\nn
    &\quad+\sum_a \gamma_h\big(-\epsilon_a-\frac{\Omega}{2}\big) \left[J_{h,a}^+ \f{\rho} J_{h,a}^- - \frac{1}{2} \left\{J_{h,a}^- J_{h,a}^+, \f{\rho}\right\}\right]\nn
    &\quad+\sum_a \gamma_h\big(+\epsilon_a+\frac{\Omega}{2}\big) \left[J_{h,a}^- \f{\rho} J_{h,a}^+ - \frac{1}{2} \left\{J_{h,a}^+ J_{h,a}^-, \f{\rho}\right\}\right]\,,
\end{align}
where $\gamma_\nu(\omega)$ are defined in~\eqref{EQ:rescorrfunc}, but are now evaluated at different transition energies, and the Lindblad jump operators are tilted
\begin{align}\label{EQ:ops_tilted}
    J_{c,a}^- &= \abs{\braket{a}{1}}^2 J_c^- + \braket{1}{a}\braket{a}{2} J_h^-\,,\nn
    J_{h,a}^- &= \abs{\braket{a}{2}}^2 J_h^- + \braket{2}{a}\braket{a}{1} J_c^-\,,
\end{align}
and $J_{\nu,a}^+ = (J_{\nu,a}^-)^\dagger$, 
where $a\in\{-,+\}$ labels the eigenstates $H_F \ket{a} = \epsilon_a \ket{a}$ of the single-particle Floquet Hamiltonian
\begin{align}\label{EQ:hamfloquet}
    H_F &= (\Delta-\Omega/2)\ket{2}\bra{2} + (\delta+\Omega/2) \ket{1}\bra{1}\nn
    &\qquad+ \lambda \ket{1}\bra{2}+ \lambda^* \ket{2}\bra{1}\,.
\end{align}
with Floquet energies
\begin{align}\label{EQ:enfloquet}
    \epsilon_\pm &= \frac{\delta+\Delta}{2} \pm \frac{1}{2}\sqrt{(\Delta-\delta-\Omega)^2+4\abs{\lambda}^2}\,.
\end{align}

The generator obtained this way is thus time-independent in the Floquet interaction picture, which facilitates the computation of stationary currents.
For the Floquet Lindblad master equation, the current leaving the reservoirs can be evaluated using counting fields, i.e., we can e.g. obtain the energy current leaving the cold reservoir via $I_E^c(t) = \ii\trace{{\cal L}'(0)\rho(t)}$, when we generalize the above Liouvillian ${\cal L} \to {\cal L}(\chi_c)$ with counting fields using the replacements (see also App.~\ref{APP:fcs} for a derivation sketch)
\begin{align}\label{EQ:replacement2}
    J_{c,a}^+ \f{\rho} J_{c,a}^- &\to J_{c,a}^+ \f{\rho} J_{c,a}^- e^{\ii(-\epsilon_a+\Omega/2)\chi_c}\,,\nn
    J_{c,a}^- \f{\rho} J_{c,a}^+ &\to J_{c,a}^- \f{\rho} J_{c,a}^+ e^{\ii(+\epsilon_a-\Omega/2)\chi_c}\,.
\end{align}
For the current leaving the cold reservoir, this yields
\begin{align}\label{EQ:current_lindblad2}
    I_E^c(t) &= \sum_a \left(\epsilon_a - \frac{\Omega}{2}\right) \Big[ \gamma_c\left(-\epsilon_a+\frac{\Omega}{2}\right) \trace{J_{c,a}^- J_{c,a}^+ \f{\rho}}\nn
    &\qquad- \gamma_c\left(\epsilon_a-\frac{\Omega}{2}\right) \trace{J_{c,a}^+ J_{c,a}^- \f{\rho}}\Big]\,.
\end{align}
The energy current leaving the hot reservoir can be treated analogously, and one obtains a similar expression (with $c\to h$ and $\Omega\to-\Omega$).

\begin{figure}
    \centering
    \includegraphics[width=0.45\textwidth]{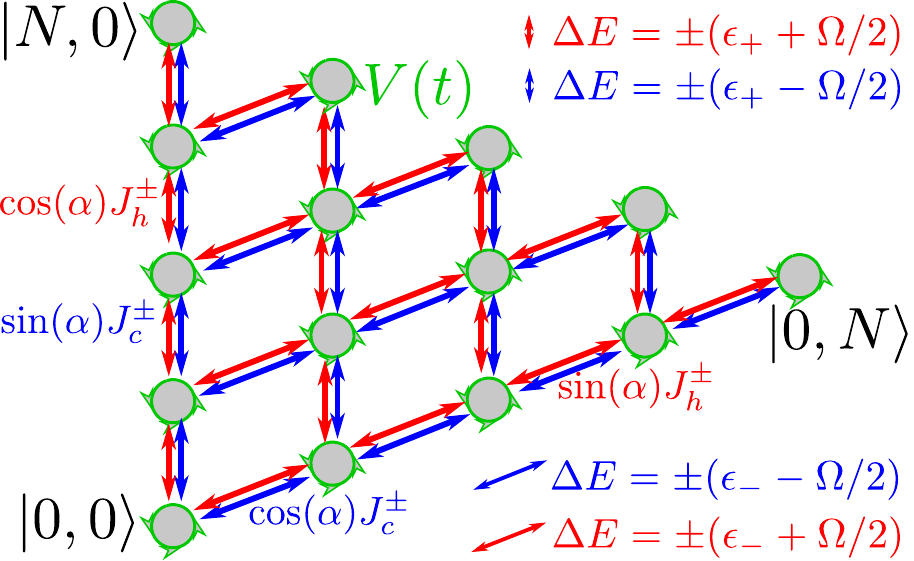}
    \caption{Sketch of the effective rate equation~\eqref{EQ:pauli_N} arising in the Floquet basis. Circles represent populations of states $\ket{M,m}$. The hot (red) and cold (blue) reservoirs induce the effective small and large transitions simultaneously. Due to the inherent periodic driving \rev{$V(t)$ (symbolized by the green arrows around circles)}, the energies exchanged with both reservoirs \rev{(indicated on the right)} are different also along the same transition, which eventually enables cooling already along the smallest cycle linking two neighboring states.}
    \label{FIG:threels4_driven}
\end{figure}

While the representation~\eqref{EQ:floquet_lindblad} provides a fixed operator basis, we show in App.~\ref{APP:floquet}
that in the collective Floquet basis constructed by (see also App.~\ref{APP:collective_bases})
\begin{align}
    \ket{M,m} \propto (S_+^+)^M (S_-^+)^m \ket{0\ldots 0}
\end{align}
with $S_a^+ = \sum_i (\ket{a}\bra{0})_i$, 
the dynamics of the diagonal matrix elements $\rho_{Mm} = \bra{M,m}\rho\ket{M,m}$ is just given by 
the Pauli-type rate equation
\begin{align}\label{EQ:pauli_N}
    \dot\rho_{M,m} &= +R_{0+}^{Mm} \rho_{M,m-1} + R_{0-}^{Mm} \rho_{M,m+1}\nn
    &\qquad+ R_{-0}^{Mm} \rho_{M+1,m} + R_{+0}^{Mm} \rho_{M-1,m} -R_{00}^{Mm} \rho_{M,m}\,,\nn
     R_{0+}^{Mm} &= \left[\gamma_c(-\epsilon_-+\frac{\Omega}{2}) \cos^2\alpha +  \gamma_h(-\epsilon_--\frac{\Omega}{2}) \sin^2\alpha\right]\times\nn
     &\qquad\times (N-M-m+1)m\,,\nn 
     R_{+0}^{Mm} &= \left[\gamma_c(-\epsilon_++\frac{\Omega}{2}) \sin^2\alpha +  \gamma_h(-\epsilon_+-\frac{\Omega}{2}) \cos^2\alpha\right]\times\nn
     &\qquad\times (N-M-m+1)M\,,\nn 
     R_{0-}^{Mm} &= \left[\gamma_c(+\epsilon_--\frac{\Omega}{2}) \cos^2\alpha + \gamma_h(+\epsilon_-+\frac{\Omega}{2}) \sin^2\alpha\right]\times\nn
     &\qquad\times (N-M-m)(m+1)\,,\nn 
     R_{-0}^{Mm} &= \left[\gamma_c(+\epsilon_+-\frac{\Omega}{2}) \sin^2\alpha + \gamma_h(+\epsilon_++\frac{\Omega}{2}) \cos^2\alpha\right]\times\nn
     &\qquad\times (N-M-m)(M+1)\,, 
\end{align}
where $R_{00}^{Mm}$ is determined by the trace conservation requirement (rate matrices must have a vanishing column sum) and $\alpha$ denotes the rotation angle of the Floquet states~\eqref{EQ:floquet_states}.
Also the counting fields from replacement~\eqref{EQ:replacement2} can be transferred to the rate equation representation (i.e., in off-diagonal matrix elements replace $\gamma_\nu(\Delta E) \to \gamma_\nu(\Delta E) e^{\ii\chi_\nu \Delta E}$).
The topology of this rate equation is depicted in Fig.~\ref{FIG:threels4_driven}.
In the absence of driving ($\lambda_i\to 0$ and $\Omega\to 0$), the eigenstates and energies fall back to the eigenstates and energies of the bare system Hamiltonian $H_S^0$, and with $\cos^2\alpha\to 1$ and $\sin^2\alpha\to 0$ one recovers the Pauli rate equation for the undriven system.
Furthermore, for resonant driving $\Omega=\Delta-\delta$ or very large amplitudes $\lambda\to\infty$, we have $\cos^2\alpha\to1/2$ and $\sin^2\alpha\to1/2$ such the prefactors can be simplified. 
In general, one can see that both reservoirs induce both effective transitions, but transferring different energies between system and hot and cold reservoir, respectively. 
This difference also implies that even at equal reservoir temperatures the stationary state will not simply be a Gibbs state of the Floquet Hamiltonian.
Apart from the significant numerical simplification, the Floquet rate equation has advantages in assessing conditions allowing for cooling functionality, as we will discuss below.
Finally, we stress that the cycles in the middle of the state network with $M\approx m\approx N/2$ have the largest Clebsch-Gordon factors in the effective Pauli rate equation~\eqref{EQ:pauli_N}, and their contribution to the cooling current scales quadratically with $N$.

\subsection{Floquet-Redfield equation}

The Floquet-Redfield master equation relies on the same interaction picture (bold symbols) as the Floquet-Lindblad master equation, with the exception that no secular approximation is performed (see App.~\ref{APP:redfield}).
Already for undriven systems, the Redfield equation~\cite{redfield1965} is in general not of Lindblad form, but for weak system-reservoir couplings it approximately preserves the system density matrix properties~\cite{hartmann2020a} and yields a thermodynamically consistent description~\cite{esposito2010b}.
Indeed, artifacts arising from the secular approximation are well-known for undriven systems (see e.g.~\cite{maekelae2013a,strasberg2018a,trushechkin2021a}) but have also been observed for driven ones~\cite{hone2009a,restrepo2018a,restrepo2019a,luo2020a}.
In the Schr\"odinger picture, the Floquet-Redfield equation can be written as 
\begin{align}\label{EQ:redfield}
    \dot\rho &= -\ii [H_S(t), \rho]\\
    &\qquad-\sum_{a\in\pm} \frac{\gamma_c(-\epsilon_a+\frac{\Omega}{2})}{2} \left\{\left[J^c, J_{c,a}^+(\Omega) \rho\right] + {\rm h.c.}\right\}\nn
    &\qquad-\sum_{a\in\pm} \frac{\gamma_c(+\epsilon_a-\frac{\Omega}{2})}{2} \left\{\left[J^c, J_{c,a}^-(\Omega) \rho\right] + {\rm h.c.}\right\}\nn
    &\qquad-\sum_{a\in\pm} \frac{\gamma_h(-\epsilon_a-\frac{\Omega}{2})}{2} \left\{\left[J^h, J_{h,a}^+(\Omega) \rho\right] + {\rm h.c.}\right\}\nn
    &\qquad-\sum_{a\in\pm} \frac{\gamma_h(+\epsilon_a+\frac{\Omega}{2})}{2} \left\{\left[J^h, J_{h,a}^-(\Omega) \rho\right] + {\rm h.c.}\right\}\,,\nonumber
\end{align}
where $\displaystyle\sum_{a\in\pm}$ again denotes the sum over the single-particle eigenstates of the Floquet Hamiltonian~\eqref{EQ:hamfloquet} but in contrast to~\eqref{EQ:ops_tilted}, the tilted operators 
\begin{align}\label{EQ:ops_tilted_rf}
    J_{c,a}^-(\Omega) &= \abs{\braket{a}{1}}^2 J_c^- + e^{+\ii\Omega t} \braket{1}{a}\braket{a}{2} J_h^-\,,\nn
    J_{h,a}^-(\Omega) &= \abs{\braket{a}{2}}^2 J_h^- + e^{-\ii\Omega t} \braket{2}{a}\braket{a}{1} J_c^-
\end{align}
with $J_{c,a}^+(\Omega) = (J_{c,a}^-(\Omega))^\dagger$ maintain a periodic time-dependence. 
In contrast to the Lindblad equations, we are not able to find a frame where the Redfield equation generator is time-independent. 
However, for our model it fortunately has only two sidebands
\begin{align}
    {\cal L}(t)={\cal L}_0+{\cal L}_- e^{-\ii\Omega t} + {\cal L}_+ e^{+\ii\Omega t}\,,
\end{align}
with the explicit superoperators given in Eq.~\eqref{EQ:sidebandsexp}.
To obtain the asymptotic solution $\bar\rho(t) = \sum_n \bar\rho^{(n)} e^{\ii n \Omega t}$ (we use the overbar generically to denote long-term limits), we thus have to solve the matrix-tridiagonal equation
\begin{align}
    0 = ({\cal L}_0 -\ii n \Omega \cdot \f{1}) \bar\rho^{(n)} + {\cal L}_+ \bar\rho^{(n-1)} + {\cal L}_- \bar\rho^{(n+1)}
\end{align}
with some suitable cutoff in $\abs{n}$ (generally chosen such that convergence is reached) for the asymptotic Fourier components $\bar\rho^{(n)}$, of which $\bar\rho^{(0)}$ provides the period-averaged density matrix.
For the parameters we investigate in detail we found a rather quick convergence with $\bar\rho^{(0)}$ being dominated by populations and $\bar\rho^{(\pm 1)}$ by coherences, and higher Fourier components vanishing rapidly with $n$.
However, we also observed that for larger qutrit numbers $N$ the necessary cutoff increased.
The currents can now be obtained by introducing counting fields (see App.~\ref{APP:fcs} for a derivation sketch) for all terms with sandwiched density matrix
\begin{align}\label{EQ:replacement3}
    J_{c,a}^-(\Omega) \rho J^c &\to J_{c,a}^-(\Omega) \rho J^c e^{\ii(\epsilon_a-\Omega/2)\chi_c}\,,\nn
    J^c \rho J_{c,a}^+(\Omega) &\to J^c \rho J_{c,a}^+(\Omega) e^{\ii(\epsilon_a-\Omega/2)\chi_c}\,,\nn
    J_{c,a}^+(\Omega) \rho J^c &\to J_{c,a}^+(\Omega) \rho J^c e^{\ii(-\epsilon_a+\Omega/2)\chi_c}\,,\nn
    J^c \rho J_{c,a}^-(\Omega) &\to J^c \rho J_{c,a}^-(\Omega) e^{\ii(-\epsilon_a+\Omega/2)\chi_c}\,,
\end{align}
which promotes ${\cal L}(t) \to {\cal L}(\chi_c,t) = {\cal L}_0(\chi_c) + {\cal L}_-(\chi_c) e^{-\ii\Omega t} + {\cal L}_+(\chi_c) e^{+\ii\Omega t}$.
The time-dependent cooling current is then obtained by $I_E^c(t) = \ii \trace{\partial_\chi {\cal L}(\chi,t)|_{\chi=0} \rho(t)}$, and it will thus not settle to a steady state value.
Therefore, to compare with the other master equations we perform a period-average~\cite{jauho1994a}.
In terms of the generalized superoperators, the period-averaged cooling current leaving the cold reservoir becomes
\begin{align}\label{EQ:pa_current_rf}
    \bar I_E^c &= (+\ii) \trace{{\cal L}_0'(0) \bar\rho^{(0)} + {\cal L}_-'(0) \bar\rho^{(+1)} + {\cal L}_+'(0) \bar\rho^{(-1)}}\,, 
\end{align}
see Eq.~\eqref{EQ:pa_current_rf_cold} for an explicit representation.
Since this only depends on the leading Fourier components $n\in\{-1,0,+1\}$, the cutoff required to obtain convergence of the current was smaller than the cutoff required for convergence of all stationary Fourier components $\bar\rho^{(n)}$.

\section{Results}\label{SEC:results}

To understand the limitations of the used master equations, we first consider the case of a single qutrit $N=1$ before continuing with the collective system.
In all our calculations, we employ a Lorentz-Drude spectral function~\cite{breuer2002} of the form
\begin{align}\label{EQ:spec_dens}
    \Gamma_\nu(\omega) = \frac{\Gamma_\nu \omega \sigma_\nu}{\sigma_\nu^2+\omega^2}
\end{align}
with bare coupling strength $\Gamma_\nu$ and width $\sigma_\nu$, 
which ensures that for all values of $\omega$, the dissipator coefficients~\eqref{EQ:rescorrfunc} remain upper-bounded.
Therefore, for small $\Gamma_\nu$, the Redfield approach is expected to be valid.
In absence of an exact solution, we thus use it as a benchmark calculation.

\subsection{Single device performance}

\subsubsection{Map of cooling}

\begin{figure}
    \includegraphics[width=0.46\textwidth]{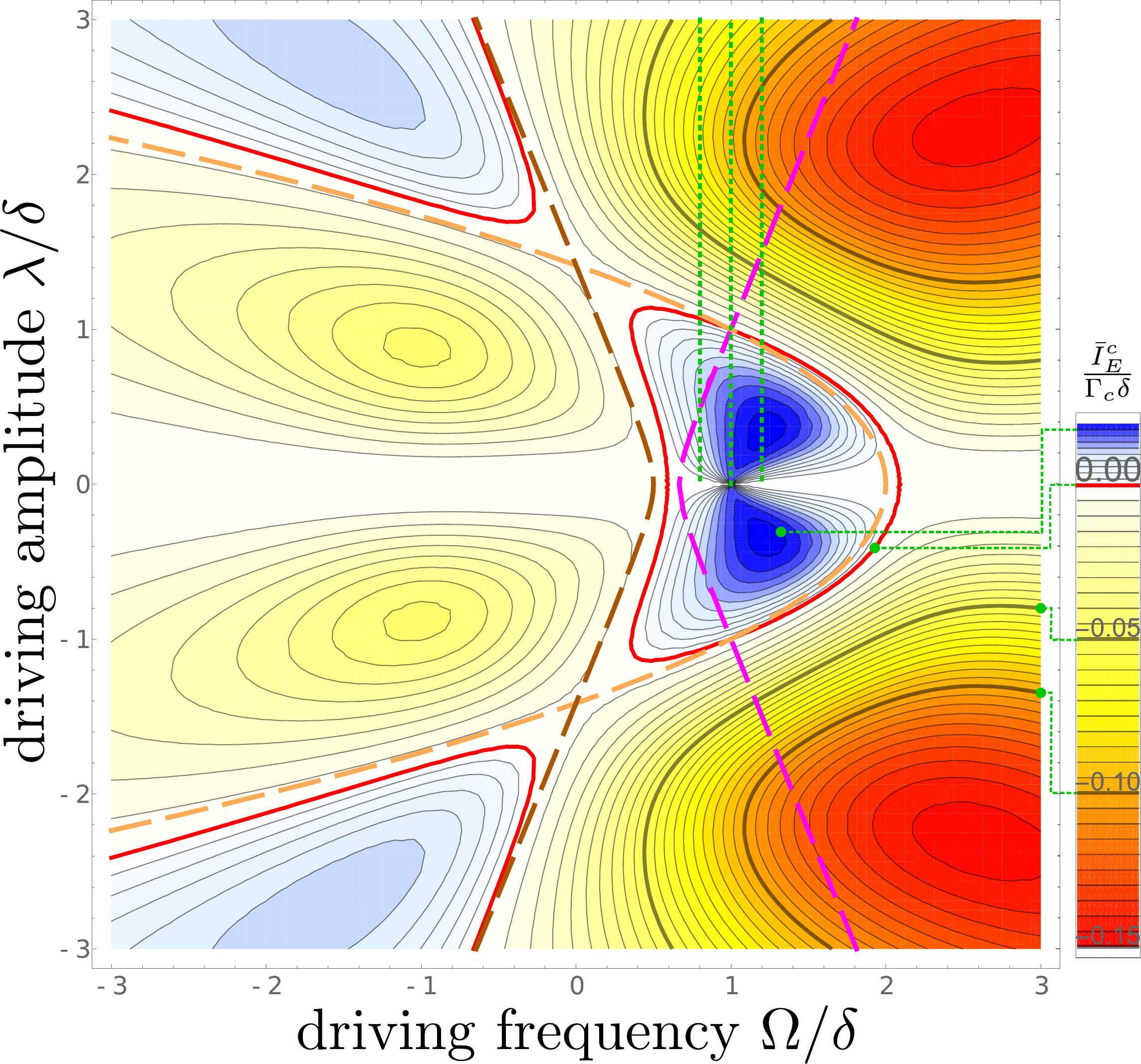}
    \caption{Plot of the stationary Floquet-Lindblad current~\eqref{EQ:current_lindblad2} versus dimensionless driving frequency $\Omega/\delta\in[-3,+3]$ (horizontal axis) and dimensionless driving strength $\lambda/\delta\in[-3,+3]$ with contours ranging from $\bar I_E^c/(\Gamma_c \delta)=-0.15$ (red) over $\bar I_E^c/(\Gamma_c \delta)=0$ (white, red contour) to $\bar I_E^c/(\Gamma_c \delta)=+0.02$ (blue). The blue regions of cooling can also be identified by cycle analysis of Pauli-type rate equation in the Floquet basis (dashed curves). Currents along the vertical green dotted lines correspond to the dashed red curves in Fig.~\ref{FIG:qutritcompcurrent}. Other parameters:
    $\Gamma_c=\Gamma_h = 0.1\delta$, $\Delta=2\delta$, $\beta_h \delta=1$, $\beta_c \delta = 1.5$, $\sigma_c=\sigma_h=\delta$.}
    \label{FIG:current_lb_1qutrit}
\end{figure}

For orientation, we first plot the stationary Floquet-Lindblad cooling current~\eqref{EQ:current_lindblad2} as a function of the driving frequency $\Omega/\delta$ and the driving strength $\lambda/\delta$ in Fig.~\ref{FIG:current_lb_1qutrit}.

First, one can see that along the vertical line defined by $\Omega=0$, the cooling current is always negative: 
In this limit, the cooling current becomes the heat flow through an undriven system, which must flow from hot to cold.

Second, along the horizontal line defined by $\lambda=0$ we see that the current vanishes -- with the exception of a small region near resonance $\lambda=0$, $\Omega=\Delta-\delta$, where the Floquet-Lindblad master equation displays artifacts.
Along this line, the current has to vanish as for $\lambda=0$, the two reservoirs induce independent cyclic transitions (compare also Fig.~\ref{FIG:qutrits_driven} in absence of the wavy line), such that no net stationary heat transfer between them is possible.

Third, the central region of cooling (dark blue colors, encircled by the solid red contour) is limited both in coupling strength (optimal finite values) and frequency (e.g. near resonance), in strong contrast to the current
obtained from the weak-driving master equation (see
App.~\ref{APP:weakdriving}). 
We can analytically provide sufficient conditions for cooling in the region bounded by the magenta dashed curve from the left and the orange dashed curve from the right, as we will explain below.

Fourth, we find two additional regions of cooling operation (top and bottom left), which for sufficiently low temperatures are found for $\Omega<0$ between the orange and brown dashed curves, as we will also explain below.

Finally, we mention that the coefficient of performance 
\begin{align}\label{EQ:cop}
    \frac{\kappa}{\kappa_{\rm Ca}} = \frac{\bar I_E^c \Theta(\bar I_E^c)}{-\bar I_E^c - \bar I_E^h} \frac{\beta_c-\beta_h}{\beta_h}
\end{align}
for the Floquet-Lindblad currents is always bounded by its Carnot value, which we demonstrate explicitly in the appendix in Fig.~\ref{FIG:cop_lb_1qutrit}.
We also stress that in contrast to the the weak-driving master equation (see App.~\ref{APP:weakdriving}), the coefficient of performance for the Floquet-Lindblad equation is not constant in the regions of cooling functionality.

\rev{Fig.~\ref{FIG:current_lb_1qutrit} already illustrates two of our main results: The first is that a microscopic Floquet treatment yields outcomes that differ qualitatively from the phenomenological approach (e.g. multiple islands of cooling functionality) and agree with the latter only in very tiny regions.
The second is that -- despite the complexity of the microscopic Floquet description -- we can find sufficient analytic conditions for cooling.
}

\subsubsection{Comparison of methods}

\begin{figure}
    \centering
    \includegraphics[width=0.49\textwidth]{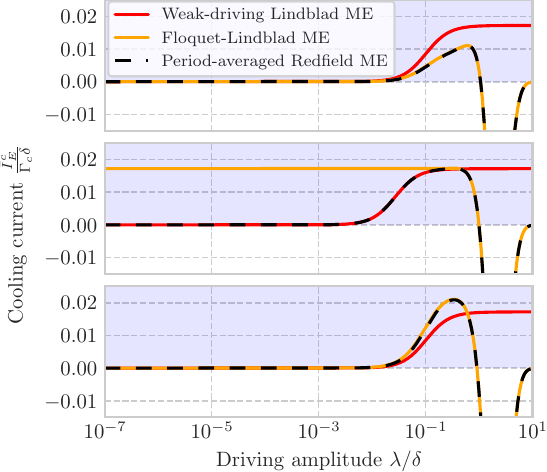}
    \caption{Plot of cooling currents for a single qutrit subject to red-detuned (top, $\Omega=0.8(\Delta-\delta)$), resonant (middle, $\Omega=\Delta-\delta$), and blue-detuned (bottom, $\Omega=1.2(\Delta-\delta)$) driving versus the dimensionless driving amplitude, i.e., along the green dotted lines in Fig.~\ref{FIG:current_lb_1qutrit}. Cooling functionality is obtained for positive currents (blue background). Lindblad master equations (MEs) only cover a finite region of validity such as small $\lambda$ (weak-driving Lindblad ME) or non-resonant driving (for resonant driving large $\lambda$) (Floquet-Lindblad ME) and may show unphysical artifacts outside: The weak-driving Lindblad ME does not capture the complete loss of cooling functionality for large $\lambda$ (right) and the Floquet Lindblad ME shows a finite cooling current at vanishing $\lambda$ for resonant driving (middle left). For the period-averaged Redfield current, Fourier modes $n\in\{-1,0,+1\}$ were sufficient for convergence. Other parameters like in Fig.~\ref{FIG:current_lb_1qutrit}.
    }
    \label{FIG:qutritcompcurrent}
\end{figure}

One may naturally ask how the map of cooling performance from Fig.~\ref{FIG:current_lb_1qutrit} compares with the results from the weak-driving master equation, which predicts cooling functionality that -- whenever $\beta_c\delta<\beta_h\Delta$ -- increases monotonically with driving strength $\lambda$ up to a maximum value (see e.g. Eqns.~\eqref{EQ:current_ad_1} and \eqref{EQ:current_ad_2} in the appendix).
In fact, analyzing this for near-resonant regions (along the dotted green lines in Fig.~\ref{FIG:current_lb_1qutrit}), we see in Fig.~\ref{FIG:qutritcompcurrent} that the weak-driving master equation results are only reproduced in the limit of small $\lambda$, where their application is justified.
Outside this region, the full Floquet-Lindblad treatment shows a turnover and for large driving amplitudes $\lambda$ eventually leads to a loss of cooling functionality.
As confirmed by analysis of the period-averaged Redfield current (black dashed curve), the Floquet-Lindblad treatment is more applicable to the regime of non-vanishing $\lambda$.
Quite analogous to the discussion of local and global master equations for undriven systems~\cite{hofer2017a}, we also stress that one should not always favor the Floquet-Lindblad master equation (''global'') over the weak-driving (''local'') one: 
For small driving strength $\lambda\approx 0$ and near resonance $\Omega\approx \Delta-\delta$ (middle panel), the Floquet-Lindblad master equation predicts a non-vanishing current where it should actually vanish.
This exception is an artifact of the Floquet-Lindblad master equation: The Floquet energies~\eqref{EQ:enfloquet} become degenerate at this point, such that the performed secular approximation is not applicable there.
The period-averaged Redfield current (black dashed curve) does not suffer from this artifact and (correctly) agrees with the weak-driving master equation in this regime.
Apart from a small region near $\Omega\approx \Delta-\delta$ and $\lambda\approx 0$ (of which we provide a comparison in Fig.~\ref{FIG:current_rf_1qutrit} in the appendix), Floquet-Redfield and Floquet-Lindblad currents agree well in the cooling current. 
We conjecture that this surprisingly good agreement is due to the chosen circular driving, which leads to a Redfield generator with just two sidebands also in the microscopic treatment.

\subsubsection{Cooling conditions for the Floquet-Lindblad equation}

\begin{figure}
    \begin{tabular}{cc}
    \includegraphics[width=0.25\textwidth]{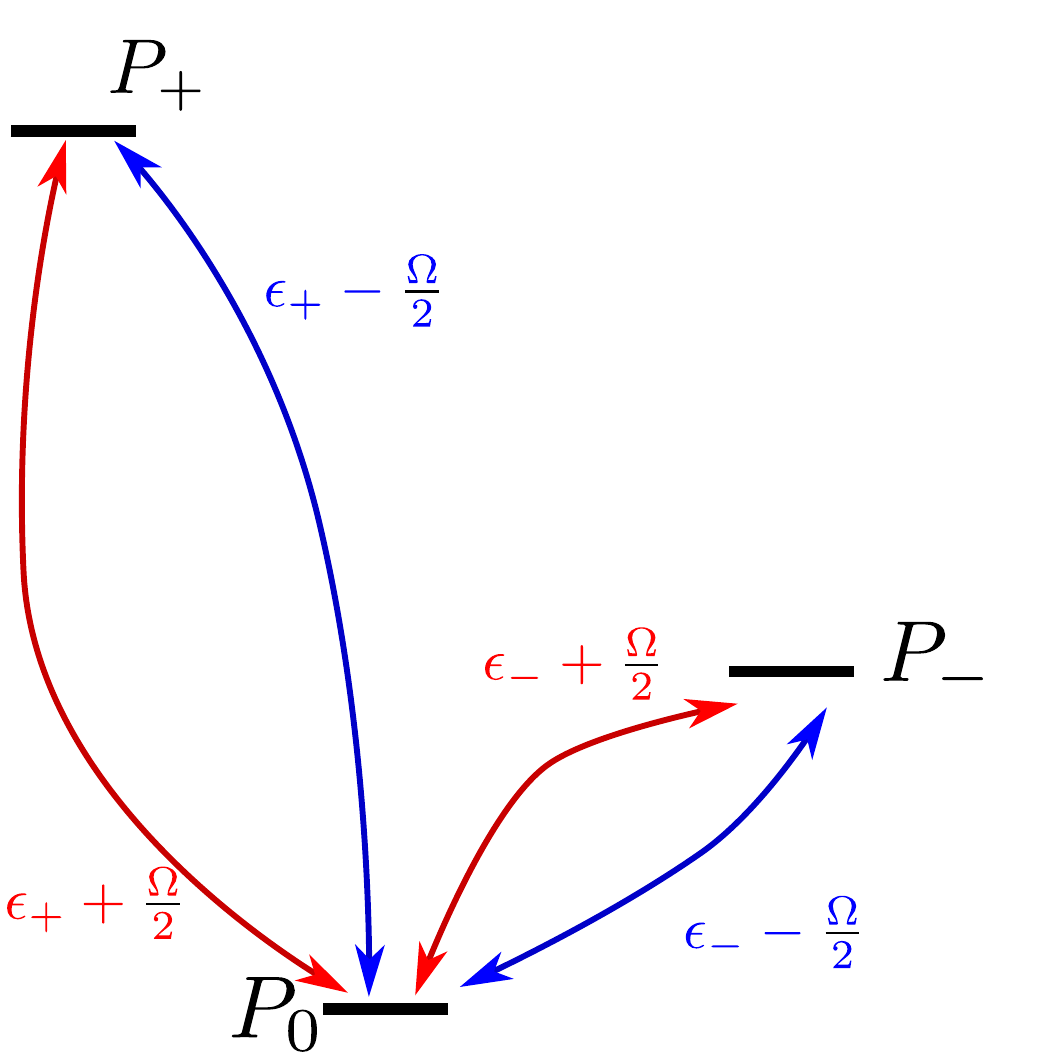} &
    \includegraphics[width=0.25\textwidth]{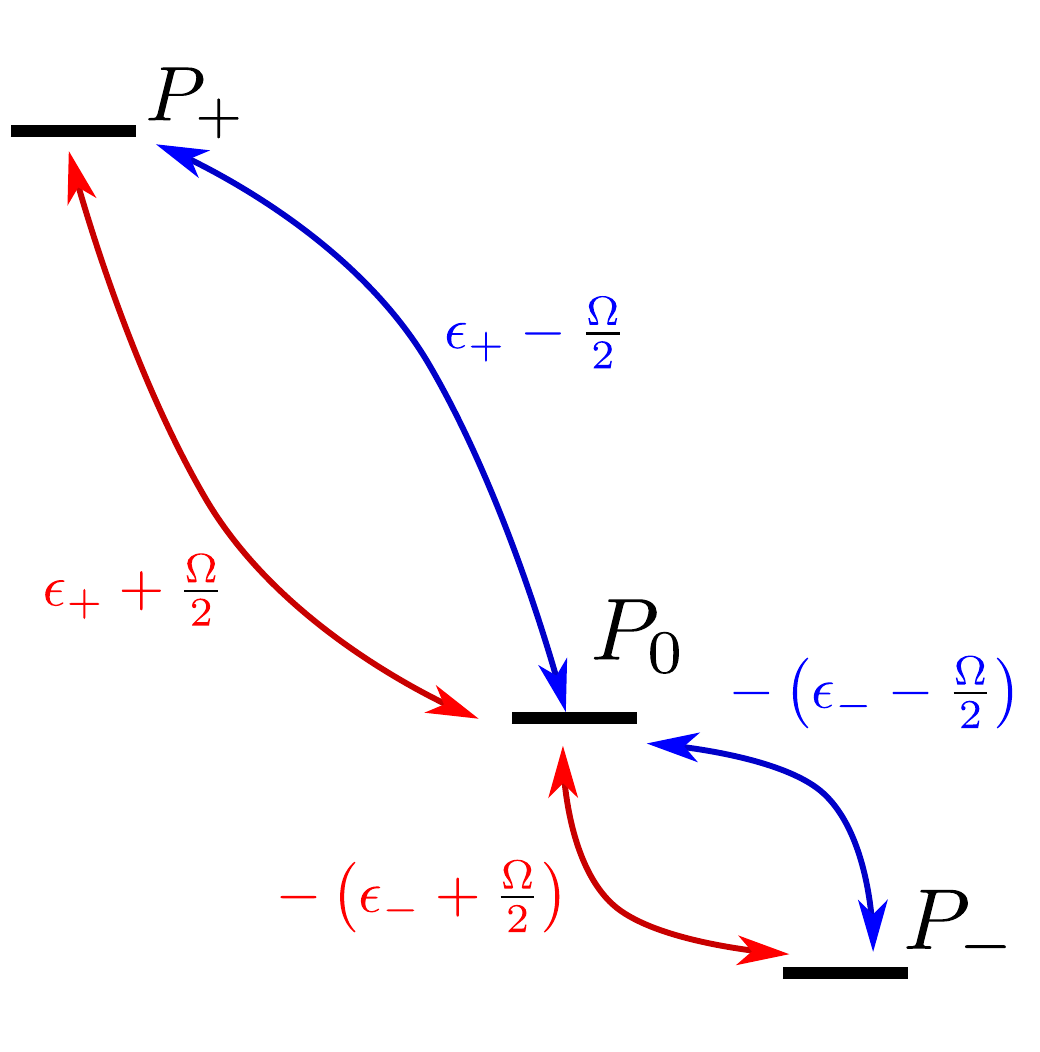}
    \end{tabular}
    \caption{Sketch of level configurations for $\Omega>0$ and $\epsilon_--\Omega/2>0$ (left panel) and for $\Omega<0$ and $\epsilon_--\Omega/2<0$ (right panel) with effective excitation energies seen by the cold (blue) or the hot (red) reservoir.
    For the left configuration, both cycles are cooling-operational (traversed counter-clockwise) when $\beta_c (\epsilon_+-\Omega/2) < \beta_h (\epsilon_++\Omega/2)$ holds.
    For the right configuration, only the lower cycle becomes operational when $\beta_c (\epsilon_--\Omega/2) > \beta_h (\epsilon_-+\Omega/2)$ (the upper cycle is then always counter-operational, traversed clockwise).
    }
    \label{FIG:coolingconfig}
\end{figure}

For $N=1$, the Floquet-Pauli master equation~\eqref{EQ:pauli_N} reduces to a three-dimensional rate equation for the probabilities $P_0=\rho_{0,0}$, $P_-=\rho_{0,1}$, and $P_+=\rho_{1,0}$ of finding the qutrit in states $\ket{0}$, $\ket{-}$ and $\ket{+}$, respectively, see Eq.~\eqref{EQ:pauli_1} in the appendix.
As a consequence of the periodic driving, it has shifted detailed-balance relations (that already for a single reservoir prohibit thermalization~\cite{shirai2015a,shirai2016a}).
Additionally, both reservoirs may now trigger both effective transitions, such that already for $N=1$ two cycles between the states $\ket{0}\leftrightarrow\ket{-}$ and $\ket{0}\leftrightarrow\ket{+}$ emerge.
These transitions are associated with reservoir-specific energetic exchanges:
The hot reservoir transfers energy $\epsilon_\pm + \Omega/2$, whereas the cold reservoir transfers energy $\epsilon_\pm - \Omega/2$.
We can analyze its cycles~\cite{schnakenberg1976a} to find conditions under which cooling performance is guaranteed.
In particular, for $\Omega>0$ and both $\epsilon_\pm-\Omega/2>0$, we have a configuration as depicted in the left panel of Fig.~\ref{FIG:coolingconfig}.
Then, a sufficient condition for cooling functionality is when both cycles are more likely traversed in a counterclockwise than in a clockwise fashion, such that from the product of rates belonging to these trajectories we obtain the conditions $\gamma_c(-\epsilon_\pm + \Omega/2) \gamma_h (\epsilon_\pm + \Omega/2) > \gamma_h(-\epsilon_\pm-\Omega/2) \gamma_c(\epsilon_\pm-\Omega/2)$.
Using that $\epsilon_+ > \epsilon_-$ and $\beta_c > \beta_h$, we see that the condition by the upper cycle is tighter, such that we can summarize the sufficient cooling conditions as
\begin{align}\label{EQ:cooling_condition}
    \epsilon_- - \frac{\Omega}{2} > 0\,,\qquad
    \beta_c (\epsilon_+-\Omega/2) < \beta_h (\epsilon_++\Omega/2)\,.
\end{align}
Given that identifying cooling conditions is already non-trivial for undriven systems~\cite{friedman2019a}, we consider this as our first main (non-technical) result. 
The above cooling condition is well supported by our numerical findings in Fig.~\ref{FIG:current_lb_1qutrit}, where the first condition denotes the region left of the orange-dashed parabola, and the second the region right of the dashed magenta curve, and indeed the cooling current is always positive when both conditions are fulfilled. 
However, from comparing with the red contour in Fig.~\ref{FIG:current_lb_1qutrit} we see that the conditions are not necessary ones as e.g. cooling performance of one of the cycles may suffice.
If we increase $\lambda$ strongly, at some point the lower Floquet energy \rev{falls significantly below zero and for all reservoirs $\beta_\nu \epsilon_- \ll -1$. Then}, essentially only the state $\ket{-}$ is occupied: The system cannot cool and even the current vanishes completely, which is what we see after the turnover of the Floquet curves in Fig.~\ref{FIG:qutritcompcurrent}. 

The functionality of the other regions for $\Omega<0$ and larger $\abs{\lambda}$ can also be understood.
Then, we have a configuration where $\epsilon_--\Omega/2<0$ (above/right/below the orange-dashed parabola in Fig.~\ref{FIG:current_lb_1qutrit}).
The condition for the lower cycle $P_- \leftrightarrow P_0$ to cool is then derived from similar arguments as 
$\beta_c(\epsilon_--\Omega/2)>\beta_h(\epsilon_-+\Omega/2)$ (left of orange dashed curve).
As this is not a sufficient condition (the other cycle is then counter-operational), the area enclosed by the dashed orange and dashed brown curves in top and bottom left parts is larger than the cooling region (red contour).
Furthermore, this operational window will depend not only on the ratio of the two reservoir temperatures, but also on the temperatures themselves.
For example, in contrast to the central cooling window, these operational windows will vanish when the temperatures of both reservoirs are increased while their ratio is kept (which is what we observed when plotting Fig.~\ref{FIG:current_lb_1qutrit} for higher temperatures, not shown).

\subsection{Collective effects}

When analyzing collective effects, we are particularly interested in enhancements of the cooling current.

\subsubsection{Cooling conditions for many qutrits}

When $N>1$ qutrits are collectively coupled to the reservoirs and are also driven collectively, the inherent permutational symmetry of the Hamiltonian preserves the symmetry of an initial condition.
Mathematically, this results in the conservation of the Casimir operators of the $su(3)$.
The subspace of complete permutational symmetry corresponds to the subspace with maximum (quadratic) Casimir operator eigenvalue, and has dimension of $(N+1)(N+2)/2$, which is much less than the $3^N$ states required for a description without permutational symmetry.
Representing all operators within a collective, completely permutationally symmetric basis (see App.~\ref{APP:collective_bases}), we show in App.~\ref{APP:weakdriving} that in the large $N$-limit, the weak-driving master equation does not enhance cooling functionality.
For finite $N$ though, we numerically find a super-linear scaling. 
For the Floquet-Lindblad master equation we find in full analogy to the $N=1$ case a Pauli-type rate equation for the populations of the density matrix in the maximum symmetry sector~\eqref{EQ:pauli_N}, which has analogous effective energy differences in its cycles. 
Therefore, our sufficient conditions~\eqref{EQ:cooling_condition} remain just the same: When all cycles cool individually, the collective system cools as well.
In contrast, if only a fraction of the cycles is cooling operational (as in Fig.~\ref{FIG:coolingconfig} right panel), cooling performance may depend on the number of qutrits involved, which is also what we observe (not shown).

\subsubsection{Collective enhancements}

Similar to Ref.~\cite{kolisnyk2023a} for undriven qutrits, we find that performance enhancements happen when the states with a number of excitations that is about half the number of qutrits (those in the central part of Fig.~\ref{FIG:threels4_driven}) are occupied.
This can be understood as follows: In the corresponding Floquet rate equation the transition rates between these states are significantly enhanced by the Clebsch-Gordan factors~\eqref{EQ:collective_matrix_element}, which also enhances the cooling current.

\begin{figure}
    \centering
    \includegraphics[width=0.5\textwidth]{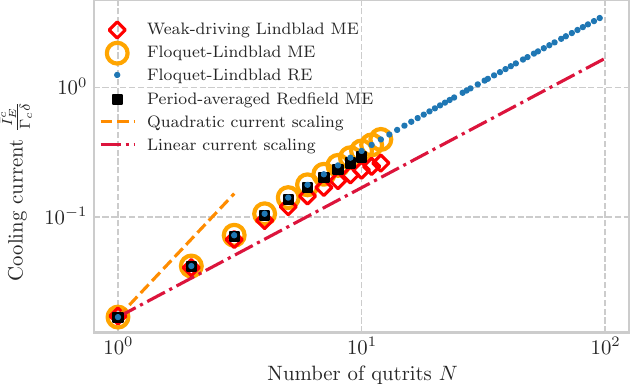}
    \caption{Multi-qutrit cooling current versus $N$ under resonant moderate strength driving $\Omega=\Delta-\delta$ and $\lambda=0.5\delta$ for different approaches (symbols). 
    The weak-driving cooling current (red diamonds) derived from~\eqref{EQ:lindblad_adiabatic} is smaller than the period-averaged Redfield cooling current (black squares) from~\eqref{EQ:redfield}, the latter is much closer to the Floquet-Lindblad cooling current (orange circles) from~\eqref{EQ:floquet_lindblad} and the Floquet-Pauli cooling current (blue dots) from the rate equation (RE) in Eq.~\eqref{EQ:pauli_N} (the last two agree exactly).
    Although the collective working fluid in all approaches (symbols) outperforms $N$ individual ones (classical scaling, red dash-dotted curve) for the sizes $N$ accessible to us, the current never scales quadratically (dashed orange line). To the contrary, in the Floquet approaches the initial superlinear scaling quickly becomes linear, and \rev{-- as an artifact of this approach in the moderate driving-strength-regime --} the scaling of the weak-driving cooling current even becomes sub-linear. For the period-averaged Redfield current, a cutoff $n_c=1$ was sufficient (the relative error $\frac{|\bar{I}^c_E(n_c=1)-\bar{I}^c_E(n_c=2)|}{\bar{I}^c_E(n_c=1)}$ was below 0.1\% for the values of $N$ that were used for plotting). Other parameters are the same as in Fig.~\ref{FIG:current_lb_1qutrit}.
    }
    \label{FIG:current_vs_n_low_temp}
\end{figure}

\begin{figure}
    \centering
    \includegraphics[width=0.5\textwidth]{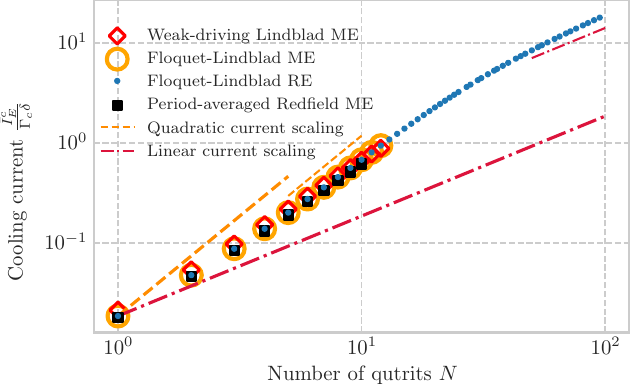}
    \caption{Analogous to Fig.~\ref{FIG:current_vs_n_low_temp}, but for higher temperatures $\beta_c \delta = 0.3$, $\beta_h \delta=0.2$. We find for small to moderate $N$ a region of quadratic scaling (thin dashed orange line shown to guide the eye), but for larger $N$ (accessible only with the Floquet-Pauli rate equation) it reduces to linear (dash-dotted red line shown to guide the eye).
    }
    \label{FIG:current_vs_n_high_temp}
\end{figure}

The population of these states can for example be achieved by using sufficiently large temperatures (while keeping their ratio constant, such that~\eqref{EQ:cooling_condition} remains fulfilled).
Indeed, we see for rather low temperatures a linear scaling with the number of qutrits (see Fig.~\ref{FIG:current_vs_n_low_temp}), whereas for higher temperatures the scaling is first quadratic (which for our type of model is the maximum possible scaling~\cite{kamimura2023a}) and turns down to a linear scaling for larger $N$ (see Fig.~\ref{FIG:current_vs_n_high_temp}).
Keeping the temperatures finite and extrapolating $N$ to infinity, we always expect a cooling performance with for large $N$ linear scaling with $N$ in the optimal cooling window defined by~\eqref{EQ:cooling_condition}.

Note however, that for the weak-driving Lindblad equation (red diamond symbols) we would even find a reduction of the cooling current for large $N$ and any finite temperatures as compared to the single-engine performance, see App.~\ref{APP:weakdriving}, and a precursor of this is already visible in Fig.~\ref{FIG:current_vs_n_low_temp}.

\rev{Whether the scaling is quadratic or linear, Figs.~\ref{FIG:current_vs_n_low_temp} and~\ref{FIG:current_vs_n_high_temp} both illustrate our third main result: The collective cooling output (symbols) may be more than the sum of its parts (red dash-dotted line).}

\section{Conclusions and Outlook}

We analyzed the cooling performance of an $N$-qutrit working fluid that is collectively coupled to two reservoirs and circularly driven as a function of driving frequency, driving strength, and qutrit number.
\rev{
Our first main result is that a full microscopic Floquet treatment leads to a completely different landscape of cooling functionality as compared to the widely-used phenomenologic approach that inserts periodic driving a posteriori in the evolution equations.
The existence of analytic sufficient conditions for cooling functionality based on a cycle analysis is our second main result.
Finally, our third main result is that superradiant enhancements can be achieved also for periodically driven systems at steady state.
}

\rev{Our first result generally poses a warning against the use of phenomenologic periodic master equation without explicit microscopic derivation that highlights its limits. 
With respect to the refrigeration ability of our system, it shows that region of cooling functionality according to the Floquet master equations is limited both in driving strength and frequency already for a single qutrit working medium.
To analyze the impact of the secular approximation, we corroborated these results with the Floquet-Redfield equation, where for our system we found Floquet-Redfield and Floquet-Lindblad approaches to yield equivalent results -- except at resonant driving $\Omega\approx \Delta-\delta$ and weak driving strength $\lambda\approx 0$. 
At this delicate spot, the secular approximation fails and the weak-driving master equation did produce better results.
We suspect that the presence of only a single delicate spot is due to the fact that the Floquet-Redfield generator for our model has only three non-vanishing Fourier components, which strongly limits the number of situations where the secular approximation fails.
For non-circular drivings like e.g. $V(t) = \cos(\Omega t) (J_w^+ + J_w^-)$ we therefore do not expect such a good agreement (compare e.g. Refs~\cite{kenmoe2016a,konghambut2021a} for driven three-level systems).
For these, the Floquet-Redfield master equation may in principle remain the method of choice~\cite{schnell2020a}, although it would certainly be interesting to analyze Floquet versions of non-secular Lindblad master equations~\cite{kirsanskas2018a,ptaszynski2019a,kleinherbers2020a,nathan2020a}. 
}

\rev{Regarding our second main result, we expect that cycle analyses of the Floquet Lindblad master equations may become a useful tool to assess useful operational modes in general, since rate equations in the Floquet basis should emerge generically.
We also remark here that it is important to correctly account for the heat exchanges with the reservoir in such effective rate equations, as the energy differences of the system alone do not contain that information.
}

\rev{For our third result, we found that -- although the scaling reduces to linear above a temperature-dependent size of the working fluid -- the overall performance of the collective device was in appropriate regimes still larger than that of independently working refrigerators.
We found in the optimal cooling regime that for fixed temperatures there is a crossover behaviour as a function of qutrit number $N$: For small $N$, the cooling performance is collectively enhanced to a quadratic scaling, since the system can populate states with collectively-enhanced transition rates.}
As $N$ grows, the finite reservoir temperatures can no longer populate these states, and the scaling is reduced to linear.
Consistent with this observation we find that at higher temperatures of both reservoirs, the transition from quadratic to linear scaling occurs at larger $N$.
We expect that fine-tuned inter-qutrit interactions (analogous to Ref.~\cite{kloc2021a}) can in principle be used to maintain the quadratic scaling also for larger $N$.
Although the Floquet treatment of interacting systems is challenging~\cite{hartmann2017a,eckardt2017a,weidinger2017a,herrmann2017a,peng2021a,rubio_abadal2020a}, the exploration of such models may yield interesting new physics~\cite{chinzei2020a}.
Departing from the idealized assumption of collective couplings and driving, we expect a quick breakdown of the quantum enhancements, as is known for undriven systems~\cite{kolisnyk2023a}.
In this case, the permutational symmetry is no longer preserved and we cannot restrict our considerations to the perfectly symmetric subspace.

\rev{On the technical side, we remark that to}
evaluate (period-averaged) energy currents \rev{leaving the reservoirs}, we used a counting field formalism, which can be combined with existing master equation approaches in a straightforward way.
\rev{Recent proposals even extend this formalism to the coherent driving field~\cite{engelhardt2024a}.}
In addition, the formalism can also be used to calculate current fluctuations~\cite{wu2010a,benito2016b,restrepo2019a}, such that thermodynamic uncertainty relations~\cite{barato2015a} (relevant e.g. for engine reliability)
can be studied.
For the long-term fluctuations $\bar S_E^\nu = \lim\limits_{t\to\infty} \frac{d}{dt} [\expval{(H_B^{\nu})^2}-\expval{H_B^\nu}^2]$ of the energy currents of bath $\nu$, the standard version of these relations predicts that
\begin{align}\label{EQ:tur}
    \frac{\bar S_E^\nu}{(\bar I_E^\nu)^2} \dot{\bar{\sigma}} \ge 2\,,
\end{align}
where $\dot{\bar{\sigma}}=-\beta_c \bar I_E^c - \beta_h \bar I_E^h \ge 0$ is the long-term entropy production rate.
For dissipative systems with periodic driving inserted a posteriori (equivalent to our weak-driving Lindblad master equation), violations of the above standard bound have been predicted~\cite{barato2018a,koyuk2019a} and observed~\cite{menczel2021a,kalaee2021b,jaseem2023a}.
While we can confirm such violations of the standard bound~\eqref{EQ:tur} for the weak-driving master equation at non-vanishing driving amplitudes $\lambda$, we do numerically not find them when using the Floquet-Lindblad approach (where we start from a periodically driven Hamiltonian).
This shows that the investigation of such relations beyond phenomenological or weak-driving approaches is an interesting future option.
\rev{Finally, we} also remark that our methods are limited to driving frequencies that are small in comparison to the reservoir relaxation timescale. 
The study of such superfast periodic drivings~\cite{das2020a} would require to go beyond a Markovian Floquet description~\cite{hotz2021a}.
%

\begin{acknowledgments}
Financial support by the DFG (project ID 278162697 -- SFB 1242) is gratefully acknowledged.    
\end{acknowledgments}

\appendix

\section{Weak-driving master equation}\label{APP:weakdriving}

\rev{In this appendix, we derive -- specifically for our system -- a master equation that is perturbative in the driving strength $\lambda$ and the system-reservoir coupling $B^\nu$. 
We recover the widely-used phenomenologic master equation that one obtains by deriving the dissipators for the undriven system and a posteriori adding the driving term to the system Hamiltonian.
We discuss the steady-state solution and general thermodynamic features and also address the limits of a single qutrit $N=1$ and the large-$N$ limit.}

\subsection{Derivation}

The derivation of the weak-driving master equation can be performed in analogy to the derivation of a local master equation for undriven systems.
That is, although $V(t)$ acts solely in the Hilbert space of the system, we consider it as a perturbation.
In the interaction picture with respect to $H_S^0+H_B^c+H_B^h$ (also denoted by bold symbols in this section), the Hamiltonian reads
\begin{align}
    \f{H}(t) &= \f{V}(t) + \sum_\nu \f{J^\nu}(t) \otimes \f{B^\nu}(t)\,,\nn
    \f{J^c}(t) &= \f{J_c^-}(t) + \f{J_c^+}(t)\,,\qquad \f{J_c^\pm}(t) = e^{\pm\ii\delta t} J_c^\pm\,,\nn 
    \f{J^h}(t) &= \f{J_h^-}(t) + \f{J_h^+}(t)\,,\qquad \f{J_h^\pm}(t) = e^{\pm\ii\Delta t} J_h^\pm\,,\nn
\f{V}(t) &= \lambda e^{+\ii(\Omega-\Delta+\delta) t} J^w_- + \lambda e^{-\ii (\Omega-\Delta+\delta)t} J^w_+\,,
\end{align}
and it enters the von-Neumann equation as $\dot{\f{\rho}}_{\rm tot} = -\ii \left[\f{H}(t), \f{\rho}_{\rm tot}(t)\right]$.
Formally integrating the von-Neumann equation and re-inserting the result
\begin{align}
    \f{\rho}_{\rm tot}(t) = \rho_S^0 \otimes \bar\rho_B -\ii \int_0^t \left[\f{H}(t'), \f{\rho}_{\rm tot}(t')\right] dt'
\end{align}
only -- \rev{this differs from the Floquet-Lindblad master equation exposed in App.~\ref{APP:floquet} but is rather analogous to the derivation of a local master equation for undriven systems~\cite{landi2022a,schaller2022a}}
-- into the terms with the reservoir coupling yields
\begin{align}
    \dot{\f{\rho}}_{\rm tot} &= -\ii \left[\f{V}(t), \f{\rho}_{\rm tot}(t)\right]\nn
    &\qquad-\ii \left[\sum_\nu \f{J^\nu}(t) \otimes \f{B^\nu}(t), \rho_S^0 \otimes \bar\rho_B\right]\nn
    &\qquad-\int_0^t dt' \left[\sum_\nu \f{J^\nu}(t) \otimes \f{B^\nu}(t), \left[\f{V}(t'), \f{\rho}_{\rm tot}(t')\right]\right]\nn
    &\qquad-\int_0^t dt' \Big[\sum_\nu \f{J^\nu}(t) \otimes \f{B^\nu}(t), \nn
    &\qquad\qquad\left[\sum_\mu \f{J^\mu}(t')\otimes\f{B^\mu}(t'), \f{\rho}_{\rm tot}(t')\right]\Big]\,.
\end{align}
Now, under the Born approximation $\f{\rho}_{\rm tot}(t) \approx \f{\rho}(t) \otimes \bar\rho_B +\ord\{\lambda\}+\ord\{\rev{B^\nu}\}$ with $\traceB{B^\nu \bar\rho_B}=0$, we can perform the partial trace and get a closed non-Markovian master equation for the system density matrix $\f{\rho}(t)$ in the interaction picture
\begin{align}
    \dot{\f{\rho}} &= -\ii \left[\f{V}(t), \f{\rho}(t)\right]-\int_0^t dt' \sum_\nu \times\nn
    &\qquad\times\traceB{\left[\f{J^\nu}(t) \f{B^\nu}(t), 
    \left[\f{J^\nu}(t')\f{B^\nu}(t'), \f{\rho}(t')\bar\rho_B\right]\right]}\nn
    &\qquad+\ord\{\lambda^2,\lambda \rev{B^\nu, (B^\nu)^3}\}\nn
    &= -\ii \left[\f{V}(t), \f{\rho}(t)\right]\nn
    &\qquad-\sum_\nu \int_0^t dt' C_\nu(t-t') \left[\f{J^\nu}(t), \f{J^\nu}(t')\f{\rho}(t')\right]\nn
    &\qquad-\sum_\nu \int_0^t dt' C_\nu(t'-t) \left[\f{\rho}(t') \f{J^\nu}(t'), \f{J^\nu}(t)\right]\,,\nn
    &\qquad+\ord\{\lambda^2,\lambda \rev{B^\nu, (B^\nu)^3}\}\nn
    &= -\ii \left[\f{V}(t), \f{\rho}(t)\right]\nn
    &\qquad-\sum_\nu \int_0^t d\tau C_\nu(+\tau) \left[\f{J^\nu}(t), \f{J^\nu}(t-\tau)\f{\rho}(t-\tau)\right]\nn
    &\qquad-\sum_\nu \int_0^t d\tau C_\nu(-\tau) \left[\f{\rho}(t-\tau) \f{J^\nu}(t-\tau), \f{J^\nu}(t)\right]\nn
    &\qquad+\ord\{\lambda^2,\lambda \rev{B^\nu, (B^\nu)^3}\}\,,
\end{align}
\rev{where already in the first line we used that the reservoir correlation functions
\begin{align}\label{EQ:corrfunc}
    C_\nu(\tau) &= \trace{\f{B^\nu}(\tau) B^\nu \bar\rho_B}\,,\nn
    \bar\rho_B&=\frac{e^{-\beta_c H^{c\phantom{h}}_B}}{\trace{e^{-\beta_c H^{c\phantom{h}\hspace{-1mm}}_B}}}\otimes \frac{e^{-\beta_h H^h_B}}{\trace{e^{-\beta_h H^h_B}}}
\end{align}
become specific to the reservoirs due to the product structure of $\bar\rho_B$.
By invoking their rapid decay, we can now perform the Markov approximation,}
which corresponds to the replacement
$\f{\rho}(t-\tau)\to\f{\rho}(t)$ and $\int_0^t dt' \to \int_0^\infty dt'$.

This yields the weak-driving Redfield-II equation
\begin{align}\label{EQ:redfield1}
    \dot{\f{\rho}} &= -\ii \left[\f{V}(t), \f{\rho}(t)\right]\nn
    &\qquad-\sum_\nu \int_0^\infty d\tau C_\nu(+\tau) \left[\f{J^\nu}(t), \f{J^\nu}(t-\tau)\f{\rho}(t)\right]\nn
    &\qquad-\sum_\nu \int_0^\infty d\tau C_\nu(-\tau) \left[\f{\rho}(t) \f{J^\nu}(t-\tau), \f{J^\nu}(t)\right]\nn
    &\qquad+\ord\{\lambda^2,\lambda \rev{B^\nu, (B^\nu)^3}\}\,,
\end{align}
which for off-resonant driving $\Omega\neq \Delta-\delta$  maintains an explicit time-dependence in the Schr\"odinger picture dissipator (not shown).
We therefore remain in the interaction picture and perform a secular approximation in the dissipator by dropping all $t$-oscillatory terms
\begin{align}
    \left[\f{J^\nu}(t), \f{J^\nu}(t-\tau)\f{\rho}(t)\right] &\to 
    \left[J_\nu^-, J_\nu^+ \f{\rho}(t)\right] e^{-\ii\Omega_\nu \tau}\\
    &\qquad+ \left[J_\nu^+, J_\nu^- \f{\rho}(t)\right] e^{+\ii\Omega_\nu \tau}\,,\nonumber
\end{align}
and analogously for the hermitian conjugate term, which allows to separate the time-dependence from the operators.
Note that we keep it in the Hamiltonian term, with the implicit assumption that near resonance $\abs{\Omega-\Delta+\delta} \ll \abs{\delta},\abs{\Delta}$, the driving term $V(t)$ oscillates considerably slower.
Then, can eventually invoke the Sokhotskij-Plemelj theorem
\begin{align}\label{EQ:spt}
     \frac{1}{2\pi} \int_0^\infty e^{+\ii\omega\tau}d\tau = \frac{1}{2}\delta(\omega)+\frac{\ii}{2\pi} {\cal P} \frac{1}{\omega}
\end{align}
with Cauchy principal value ${\cal P}$, to write the half-sided integrals as
\begin{align}
    \int_0^\infty C_\nu(+\tau) e^{+\ii\Omega_\nu\tau} d\tau &= \frac{\gamma_\nu(+\Omega_\nu)}{2}+\frac{\sigma_\nu(+\Omega_\nu)}{2}\,,\nn
    \int_0^\infty C_\nu(-\tau) e^{+\ii\Omega_\nu\tau} d\tau &= \frac{\gamma_\nu(-\Omega_\nu)}{2}-\frac{\sigma_\nu(-\Omega_\nu)}{2}\,,
\end{align}
where $\gamma_\nu(\omega) = \int C_\nu(\tau) e^{+\ii\omega\tau} d\omega$ is the even Fourier transform of the reservoir correlation function and $\sigma_\nu(\omega) = \frac{\ii}{\pi} {\cal P}\int \frac{\gamma_\nu(\omega')}{\omega-\omega'} d\omega'$ generates the Lamb-shift terms.
Back in the Schr\"odinger picture, we then obtain a time-dependent Lindblad form, which by neglecting the Lamb-shift-terms is given by~\eqref{EQ:lindblad_adiabatic} in the main text.

\begin{figure}
    \includegraphics[width=0.45\textwidth]{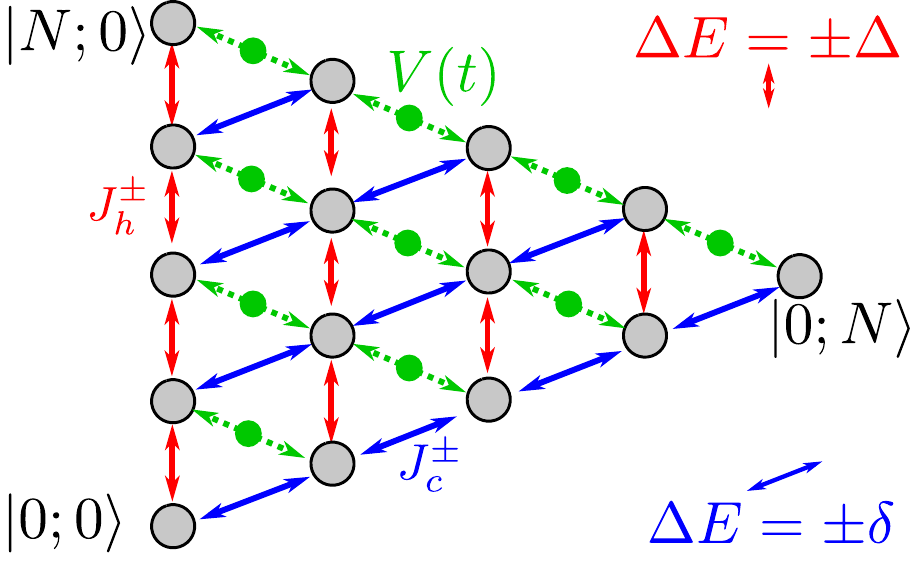}
    \caption{Visualization of incoherent and selected coherent transitions (red/blue arrows and green ones, respectively) in the weak-driving master equation~\eqref{EQ:lindblad_adiabatic}. 
    The dissipators induce transitions between the populations (red and blue arrows for hot and cold reservoir \rev{with energy differences indicated on the right}, respectively), but the \rev{driving $V(t)$} 
    couples populations in the bare system energy eigenbasis (gray spheres) to coherences (green spheres), which couple to further coherences along the green lines within the fully symmetric subspace (not shown).}
    \label{FIG:threels4_naive}
\end{figure}

The dissipative terms act like cold and hot reservoirs of a QAR, but the Hamiltonian couples populations to coherences and vice versa.
The structure of the resulting master equation within the fully symmetric subspace (see App.~\ref{APP:collective_bases}) is depicted in Fig~\ref{FIG:threels4_naive}.
The master equation is expected to be valid when both the coupling to the reservoirs is small and the driving is weak (e.g. $\lambda=\ord\{\Gamma_\nu\}$).

\subsection{Stationary frame}

With the transformation 
\begin{align}\label{EQ:rftrafo}
\rho_{\rm r}(t) &= e^{+\ii X t} \rho(t) e^{-\ii X t}\,,\nn
X &= \delta \hat{N}_\delta + (\delta+\Omega) \hat{N}_\Delta\,,
\end{align}
we find that the weak-driving master equation~\eqref{EQ:lindblad_adiabatic} becomes -- even for off-resonant driving -- time-independent in the new frame
\begin{align}\label{EQ:master_rotframe}
\dot{\rho}_{\rm r} &= -\ii \left[(\Delta-\delta-\Omega) \hat{N}_\Delta + \lambda (\hat J_w^+ + \hat J_w^-), \rho_{\rm r}\right]\nn
&\qquad+\sum_\nu \gamma_\nu(+\Omega_\nu)\left[J^\nu_- \rho_{\rm r} J^\nu_+ - \frac{1}{2}\left\{J^\nu_+ J^\nu_-, \rho_{\rm r}\right\}\right]\nn
&\qquad+\sum_\nu \gamma_\nu(-\Omega_\nu)\left[J^\nu_+ \rho_{\rm r} J^\nu_- - \frac{1}{2}\left\{J^\nu_- J^\nu_+, \rho_{\rm r}\right\}\right]\,.
\end{align}
Therefore, we expect the system to settle to a stationary state $\bar\rho_{\rm r}$ in the rotating frame, and the (asymptotically periodic) long-term state is then obtained by inverting~\eqref{EQ:rftrafo}.

\subsection{Thermodynamics}\label{SEC:thermodynamics_adiabatic}

By looking at the energy balance of $\hat H_S^0$, i.e., the bare, unperturbed system Hamiltonian, we see that it decomposes into three contributions, which we can identify as the power injected by the laser
\begin{align}
\bar P &= -\ii \lim_{t\to\infty} \trace{[\hat{H}_S^0, V(t)] \rho(t)}\nn
&= -\ii \trace{[\hat{H}_S^0, \lambda (J_w^- + J_w^+)] \bar\rho_{\rm r}}\,, 
\end{align}
and the energy currents entering the system from the cold and hot reservoirs
\begin{align}
\bar{I}_E^\nu = \lim_{t\to\infty} \trace{\hat{H}_S^0 ({\cal L}_\nu \rho(t))} = \trace{\hat{H}_S^0 ({\cal L}_\nu \bar\rho_{\rm r})}\,,
\end{align}
which all assume time-independent long-term values as can be seen by evaluating them in the rotating frame.
By construction, the first law of thermodynamics at steady state
\begin{align}\label{EQ:firstlaw}
    \bar P + \bar I_E^c + \bar I_E^h = 0
\end{align}
is automatically fulfilled.
With the definitions of the currents, and noting that ${\cal L}_\nu \bar\rho_\nu=0$ with $\bar\rho_\nu = \frac{e^{-\beta_\nu H_S^0}}{\trace{e^{-\beta_\nu H_S^0}}}$, we can also write the entropy change of the system as
\begin{align}
    \dot S &= -\trace{(\sum_\nu {\cal L}_\nu \rho) \ln \rho}\nn
    &= -\trace{(\sum_\nu {\cal L}_\nu \rho) [\ln \rho-\ln\bar\rho_\nu]}+ \sum_\nu \beta_\nu I_E^\nu\nn
    &\equiv \dot{\sigma}_\ii + \sum_\nu \beta_\nu I_E^\nu\,,
\end{align}
where by Spohn's inequality~\cite{spohn1978b,landi2022a} we can infer that the irreversible entropy production is positive $\dot{\sigma}_\ii\ge 0$.
Solving for it, and using that at steady-state $\dot{\bar{S}}= 0$, we thus obtain
\begin{align}\label{EQ:secondlaw}
    \dot{\bar{\sigma}}_\ii = -\sum_\nu \beta_\nu \bar I_E^\nu \ge 0\,,
\end{align}
which bounds the currents of the weak-driving Lindblad master equation.
This also bounds the coefficient of performance -- in regimes where the cold reservoir is cooled $\bar I_E^c > 0$ -- by its Carnot value
\begin{align}
    \kappa &= \frac{\bar I_E^c}{\bar P} = \frac{\beta_h \bar I_E^c}{(\beta_c-\beta_h) \bar I_E^c - \beta_h \bar I_E^h - \beta_c \bar I_E^c}\nn
    &\le \frac{\beta_h}{\beta_c-\beta_h} = \frac{T_c}{T_h-T_c} = \kappa_{\rm Ca}\,,
\end{align}
where we have used~\eqref{EQ:firstlaw} to eliminate $\bar P$ and~\eqref{EQ:secondlaw} to obtain the inequality.

The above definitions of the currents have been discussed for $N=1$ before~\cite{kalaee2021a}, invoking thermodynamic consistency arguments.
Here, we have recovered them from a perturbative treatment of the driving strength.
In the main text we argue that within the validity region of the weak-driving Lindblad equation, a microscopic discussion based on energy counting fields would lead to identical results for the currents leaving the reservoirs.

\subsection{Case of a single qutrit}

In the rotating frame, we get for the populations $\rho_{\rm r}^{00}$, $\rho_{\rm r}^{11}$, $\rho_{\rm r}^{22}$ and two relevant coherences $\rho_{\rm r}^{12}$ and $\rho_{\rm r}^{21}$ (the other four coherences vanish in the long-term limit, if not absent already from the beginning) the matrix representation for the Liouvillian
\begin{align}
    {\cal L} &= \left(\begin{array}{ccc|cc}
    & & & 0 & 0\\
    & {\cal L}_{\rm pop} & & +\ii\lambda & -\ii\lambda\\
    & & & -\ii \lambda & +\ii\lambda\\
    \hline
    0 & +\ii\lambda & -\ii\lambda & \xi & 0\\
    0 & -\ii\lambda & +\ii\lambda  & 0 & \xi^*
    \end{array}\right)\,,\nn
    {\cal L}_{\rm pop} &= \left(\begin{array}{ccc}
    -\gamma_c(-\delta) - \gamma_h(-\Delta) & \gamma_c(+\delta) & \gamma_h(+\Delta)\\
    \gamma_c(-\delta) & -\gamma_c(+\delta) & 0\\
    \gamma_h(-\Delta) & 0 & -\gamma_h(+\Delta)
    \end{array}\right)\,,\nn
    \xi &= -\frac{\gamma_c(+\delta)+\gamma_h(+\Delta)}{2}+\ii(\Delta-\delta-\Omega)\,,
\end{align}
where we have $\gamma_\nu(\omega)$ defined as in~\eqref{EQ:rescorrfunc} with the reservoir spectral functions analytically continued to negative frequencies as odd functions $\Gamma_\nu(-\omega)=-\Gamma_\nu(+\omega)$.
From this, e.g. steady-state solutions and cooling currents can be evaluated in a straightforward way~\cite{kalaee2021a}.
We find that the cooling current rises monotonically from zero at $\lambda=0$ to its maximum value 
(using the short-term notations $n_c = [e^{\beta_c\delta}-1]^{-1}$ and $n_h=[e^{\beta_h\Delta}-1]^{-1}$)
\begin{align}\label{EQ:current_ad_1}
    \lim_{\lambda\to\infty} \bar I_E^c = \frac{\Gamma_c(\delta) \Gamma_h(\Delta) \delta(n_c-n_h)}{\Gamma_c(\delta)(1+3n_c)+\Gamma_h(\Delta)(1+3n_h)}\,,
\end{align}
and the weak-driving master equation predicts cooling for $n_c>n_h$ (equivalently $\beta_c \delta < \beta_h \Delta$).
However, one should keep in mind that in particular the large-$\lambda$ it should not be expected to be valid.
The stationary coherence vanishes at extreme values of the driving strength $\lim_{\lambda\to 0} \abs{\bar\rho_{12}}=\lim_{\lambda\to\infty} \abs{\bar\rho_{12}} = 0$, but has a finite value with a single maximum in between.
Furthermore, we note that the currents are tightly coupled, such that the coefficient of performance is given by the constant value
$\kappa = \delta/(\Delta-\delta)$ in all regions where the cooling current is positive (such that we also always have $\kappa < \kappa_{\rm Ca}$).

\subsection{Large-\texorpdfstring{$N$}{N}-limit}

In the fully-symmetric subspace, the collective ladder operators responsible for the transitions between the states $\ket{M;m}$ can be represented by two effective bosonic modes (compare App.~\ref{APP:collective_bases})
\begin{align}
    J_h^+ &= a_\Delta^\dagger \sqrt{N-a_\Delta^\dagger a_\Delta - a_\delta^\dagger a_\delta} \approx \sqrt{N} a_\Delta^\dagger\,,\nn
    J_c^+ &= a_\delta^\dagger \sqrt{N-a_\Delta^\dagger a_\Delta - a_\delta^\dagger a_\delta} \approx \sqrt{N} a_\delta^\dagger\,,\nn
    J_w^+ &= a_\Delta^\dagger a_\delta\,,
\end{align}
which correspond to the number of small and large excitations present in the system.
From this, we see that in the many-qutrit limit $N \gg \expval{a_\Delta^\dagger a_\Delta},\expval{a_\delta^\dagger a_\delta}$, the master equation in the rotating frame~\eqref{EQ:master_rotframe} becomes quadratic in the annihilation and creation operators of these effective bosonic modes
\begin{align}
\dot{\rho}_{\rm r} &= -\ii \left[(\Delta-\delta-\Omega) a_\Delta^\dagger a_\Delta + \lambda (a_\Delta^\dagger a_\delta + a_\delta^\dagger a_\Delta), \rho_{\rm r}\right]\nn
&\qquad+N \gamma_c(+\delta)\left[a_\delta \rho_{\rm r} a_\delta^\dagger - \frac{1}{2}\left\{a_\delta^\dagger a_\delta, \rho_{\rm r}\right\}\right]\nn
&\qquad+N \gamma_c(-\delta)\left[a_\delta^\dagger \rho_{\rm r} a_\delta - \frac{1}{2}\left\{a_\delta a_\delta^\dagger , \rho_{\rm r}\right\}\right]\nn
&\qquad+N \gamma_h(+\Delta)\left[a_\Delta \rho_{\rm r} a_\Delta^\dagger - \frac{1}{2}\left\{a_\Delta^\dagger a_\Delta, \rho_{\rm r}\right\}\right]\nn
&\qquad+N \gamma_h(-\Delta)\left[a_\Delta^\dagger \rho_{\rm r} a_\Delta - \frac{1}{2}\left\{a_\Delta a_\Delta^\dagger, \rho_{\rm r}\right\}\right]\,,
\end{align}
such that it can be solved by looking at the equations of motion, which at resonance $\Omega=\Delta-\delta$ become
\begin{align}
    \partial_t \expval{a_\Delta^\dagger a_\Delta} &= -\ii \lambda\left(\expval{a_\Delta^\dagger a_\delta}-\expval{a_\delta^\dagger a_\Delta}\right)\nn
    &\qquad-N[\gamma_h(+\Delta)-\gamma_h(-\Delta)]\expval{a_\Delta^\dagger a_\Delta}\nn
    &\qquad+ N \gamma_h(-\Delta)\,,\nn
    \partial_t \expval{a_\delta^\dagger a_\delta} &= +\ii \lambda\left(\expval{a_\Delta^\dagger a_\delta}-\expval{a_\delta^\dagger a_\Delta}\right)\nn
    &\qquad-N[\gamma_c(+\delta)-\gamma_c(-\delta)]\expval{a_\delta^\dagger a_\delta}\nn
    &\qquad+ N \gamma_c(-\delta)\,,\nn
    \partial_t \expval{a_\Delta^\dagger a_\delta} &= \frac{N}{2}\Big[\gamma_c(-\delta)+\gamma_h(-\Delta)\nn
    &\qquad-\gamma_c(+\delta)-\gamma_h(+\Delta)\Big]\expval{a_\Delta^\dagger a_\delta}\nn
    &\qquad-\ii\lambda\left(\expval{a_\Delta^\dagger a_\Delta}-\expval{a_\delta^\dagger a_\delta}\right)\,,
\end{align}
and analogous for $\expval{a_\delta^\dagger a_\Delta}$.
We can solve for the steady-state expectation values and then from this evaluate the cooling current
$I_E^c = N \delta \gamma_c(-\delta) [1+\expval{a_\delta^\dagger a_\delta}] - N \delta \gamma_c(+\delta) \expval{a_\delta^\dagger a_\delta}$ in the steady state limit (we use the short-term notations $\gamma_c(+\delta)=\tilde\Gamma_c (1+n_c)$, $\gamma_c(-\delta)=\tilde\Gamma_c n_c$, $\gamma_h(+\Delta)=\tilde\Gamma_h(1+n_h)$ and $\gamma_h(-\Delta)=\tilde\Gamma_h n_h$)
\begin{align}\label{EQ:current_ad_2}
    \bar I_E^c = \frac{4 N \delta \tilde\Gamma_c \tilde\Gamma_h (n_c-n_h)\lambda^2}{(\tilde\Gamma_c+\tilde\Gamma_h)(\tilde\Gamma_c \tilde\Gamma_h N^2+4\lambda^2)}\,.
\end{align}
Unfortunately, although the cooling condition is the same as before, it seems that the cooling current only grows linearly and even decays for large $N$.
In the main text, this turnover to sub-linear scaling is already visible in Fig.~\ref{FIG:current_vs_n_low_temp} (red diamond symbols).
Even when the driving is infinitely strong, we recover at best a linear scaling.

\section{Floquet-Lindblad master equation}\label{APP:floquet}

\rev{In this appendix, we provide a derivation of the Lindblad master equation for our system that is perturbative in the system-reservoir coupling strength only. We discuss general thermodynamic features of this master equation and make formulas explicit for the case $N=1$.}

\subsection{Derivation}

The system time evolution operator can always be deomposed based on Floquet theory.
Due to the circular driving employed in our particular model, this is possible in closed form (for simplicity just for one qutrit)
\begin{align}\label{EQ:unitary_global}
    U_S(t) &= U_{\rm kick}(t) e^{-\ii H_F t}\,,\nn
    U_{\rm kick}(t) &=e^{+\ii \Omega t/2[\ket{1}\bra{1}-\ket{2}\bra{2}]} \,,
\end{align}
with Floquet-Hamiltonian given by~\eqref{EQ:hamfloquet}.
The decomposition above is not unique, we could e.g. shift the exponential in the first factor (kick operator) by $\Omega t/2 (\ket{1}\bra{1}+\ket{2}\bra{2})$ to make it periodic in the driving and modify the Floquet Hamiltonian~\eqref{EQ:hamfloquet} and its energies~\eqref{EQ:enfloquet} accordingly (which would not alter our final results).
For real-valued $\lambda$, the eigenstates can be written as
\begin{align}\label{EQ:floquet_states}
    \ket{-} &= \cos(\alpha) \ket{1} - \sin(\alpha) \ket{2}\,,\nn
    \ket{+} &= \sin(\alpha) \ket{1} + \cos(\alpha) \ket{2}\,,
\end{align}
where the rotation angle $\alpha$ obeys
\begin{align}
    \tan(\alpha) &= \frac{\Omega-\Delta+\delta+\sqrt{4\lambda^2+(\Omega-\Delta+\delta)^2}}{2\lambda}\,.
\end{align}
For weak ($\lambda\to 0$) and red-detuned ($\Omega<\Delta-\delta$) driving, we thus have $\ket{-}\to\ket{1}$ and $\ket{+}\to\ket{2}$, whereas for weak and blue-detuned driving these just exchange $\ket{-}\to-\ket{2}$ and $\ket{+}\to\ket{1}$.
In the limits of strong driving ($\lambda\to+\infty$) or also resonant driving ($\Omega=\Delta-\delta$) with finite amplitude ($\lambda>0$) we have $\epsilon_\pm \to (\Delta+\delta)/2  \pm \lambda$ and$\ket{\pm}=\frac{1}{\sqrt{2}}[\ket{1}\pm\ket{2}]$.

From the above, it follows that the collective generalization of the kick operator just provides a phase to the reservoir coupling operators.
\begin{align}
    U_{\rm kick}^{\dagger}(t) J_c^\pm U_{\rm kick}(t) &= e^{\mp\ii\Omega t/2} J_c^\pm\,,\nn
    U_{\rm kick}^{\dagger}(t) J_h^\pm U_{\rm kick}(t) &= e^{\pm\ii\Omega t/2} J_h^\pm\,.
\end{align}
With this, we now perform the derivation of the Floquet master equation. 
That is, we now use an interaction picture with a perturbative treatment of $H_I^c+H_I^h$, where the Hamiltonian reads
\begin{align}
    \f{H_I}(t) &= \sum_\nu \f{J^\nu}(t) \otimes \f{B^\nu}(t)\,,
\end{align}
where $\f{B^\nu}(t) = e^{+\ii H_B^\nu t} B^\nu e^{-\ii H_B^\nu t}$, $\f{J^\nu}(t) = U_S^\dagger(t) J^\nu U_S(t)$, and where the Hamiltonian
enters the von-Neumann equation via $\dot{\f{\rho}}_{\rm tot} = -\ii \left[\f{H}(t), \f{\rho}_{\rm tot}(t)\right]$.
After the standard steps (Born approximation using a product state of two thermal reservoirs and two Markov approximations), this leads to the Floquet-Redfield equation, analogous to Eq.~\eqref{EQ:redfield1}
\begin{align}\label{EQ:redfield2}
    \dot{\f{\rho}} &= -\sum_\nu \int_0^\infty d\tau C_\nu(+\tau) \left[\f{J^\nu}(t), \f{J^\nu}(t-\tau)\f{\rho}(t)\right]+{\rm h.c.}\,,
\end{align}
where the difference however is that the interaction picture is meant with respect to the full time-dependent system Hamiltonian $H_S(t)$, such that consistently the Hamiltonian term is missing.
\rev{The correlation functions are given as in Eq.~\eqref{EQ:corrfunc}.}

To perform a secular approximation, one can make the time-dependence explicit 
\begin{align}
    \f{J^c}(t) &= \sum_{a\in\pm} e^{-\ii(\epsilon_a-\Omega/2)t} \left[\abs{\braket{a}{1}}^2 J_c^- + \braket{1}{a}\braket{a}{2} J_h^-\right]\nn
    &\quad+ \sum_{a\in\pm} e^{+\ii(\epsilon_a-\Omega/2)t} \left[\abs{\braket{a}{1}}^2 J_c^+ + \braket{2}{a}\braket{a}{1} J_h^+\right]\nn
    &= \sum_{a\in\pm}  (e^{-\ii(\epsilon_a-\Omega/2)t} J_{c,a}^- + e^{+\ii(\epsilon_a-\Omega/2)t} J_{c,a}^+)\,,\nn
    \f{J^h}(t) &= \sum_{a\in\pm} e^{-\ii(\epsilon_a+\Omega/2)t} \left[\abs{\braket{a}{2}}^2 J_h^- + \braket{2}{a}\braket{a}{1} J_c^-\right]\nn
    &\quad+ \sum_{a\in\pm} e^{+\ii(\epsilon_a+\Omega/2)t} \left[\abs{\braket{a}{2}}^2 J_h^+ + \braket{1}{a}\braket{a}{2} J_c^+\right]\nn
    &= \sum_{a\in\pm} (e^{-\ii(\epsilon_a+\Omega/2)t} J_{h,a}^- + e^{+\ii(\epsilon_a+\Omega/2)t} J_{h,a}^+)\,.
\end{align}
When inserting the above and the analogue for $\f{J^\nu}(t-\tau)$ into the Floquet-Redfield equation, we get a double summation over Floquet energies, where one can use now that for sufficiently strong driving (e.g. at resonance) or also for moderate $\lambda$ and strong detuning, 
the Floquet energies $\epsilon_\pm$ are sufficiently distinct, such that the contributions for different $a$ are always negligible, and the ones for equal $a$ are only non-vanishing when exactly counter-rotating, leading to the Floquet-Lindblad master equation with one summation only
\begin{align}
    \dot{\f{\rho}} &= -\int_0^\infty d\tau C_c(+\tau) \sum_a e^{-\ii(\epsilon_a-\Omega/2)\tau} \left[J_{c,a}^-, J_{c,a}^+\f{\rho}\right]\nn
    &\qquad-\int_0^\infty d\tau C_c(+\tau) \sum_a e^{+\ii(\epsilon_a-\Omega/2)\tau} \left[J_{c,a}^+, J_{c,a}^-\f{\rho}\right]\nn
    &\qquad- \int_0^\infty d\tau C_h(+\tau) \sum_a e^{-\ii(\epsilon_a+\Omega/2)\tau} \left[J_{h,a}^-, J_{h,a}^+\f{\rho}\right]\nn
    &\qquad- \int_0^\infty d\tau C_h(+\tau) \sum_a e^{+\ii(\epsilon_a+\Omega/2)\tau} \left[J_{h,a}^+, J_{h,a}^-\f{\rho}\right]\nn
    &\qquad + {\rm h.c.}\,,
\end{align}
which, when we insert the Fourier transforms of the reservoir correlation functions, then use the Sokhotskij-Plemelj theorem~\eqref{EQ:spt} and neglect the Lamb-shift, directly transforms into~\eqref{EQ:floquet_lindblad} in the main text.
The above master equation can be analyzed with counting fields in a thermodynamically consistent way~\cite{bulnes_cuetara2015a}.
Further, some rewriting shows that for $a\in\pm$ the tilted collective operators actually mediate transitions between eigenstates of the collective Floquet Hamiltonian, that are just rotated compared to the original collective eigenstates
\begin{align}
    J_{c,a}^- &= \braket{1}{a} \sum_i (\ket{0}\bra{a})_i \equiv \braket{1}{a} S_a^-\,,\nn
    J_{h,a}^- &= \braket{2}{a} \sum_i (\ket{0}\bra{a})_i \equiv \braket{2}{a} S_a^-\,,
\end{align}
and analogous for $J_{\nu,a}^+$.
We can thus alternatively write the Floquet Lindblad equation~\eqref{EQ:floquet_lindblad} with raising and lowering operators between collective Floquet eigenstates
\begin{align}\label{EQ:floquet_lindblad1}
        \dot{\f{\rho}} &=\sum_a \left[\gamma_c(-\epsilon_a+\frac{\Omega}{2}) \abs{\braket{a}{1}}^2 +  \gamma_h(-\epsilon_a-\frac{\Omega}{2}) \abs{\braket{a}{2}}^2 \right]\times\nn
        &\qquad\times\left[S_a^+ \f{\rho} S_a^- - \frac{1}{2} \left\{S_a^- S_a^+, \f{\rho}\right\}\right]\nn
    &+\sum_a \left[\gamma_c(+\epsilon_a-\frac{\Omega}{2}) \abs{\braket{a}{1}}^2 + \gamma_h(+\epsilon_a+\frac{\Omega}{2}) \abs{\braket{a}{2}}^2\right]\times\nn
    &\qquad\times\left[S_a^- \f{\rho} S_a^+ - \frac{1}{2} \left\{S_a^+ S_a^-, \f{\rho}\right\}\right]\,.
\end{align}
Evaluating this equation in the energy eigenbasis of the Floquet Hamiltonian leads to Eq.~\eqref{EQ:pauli_N} in the main text.

\subsection{Thermodynamic discussion}

\begin{figure}
    \includegraphics[width=0.46\textwidth]{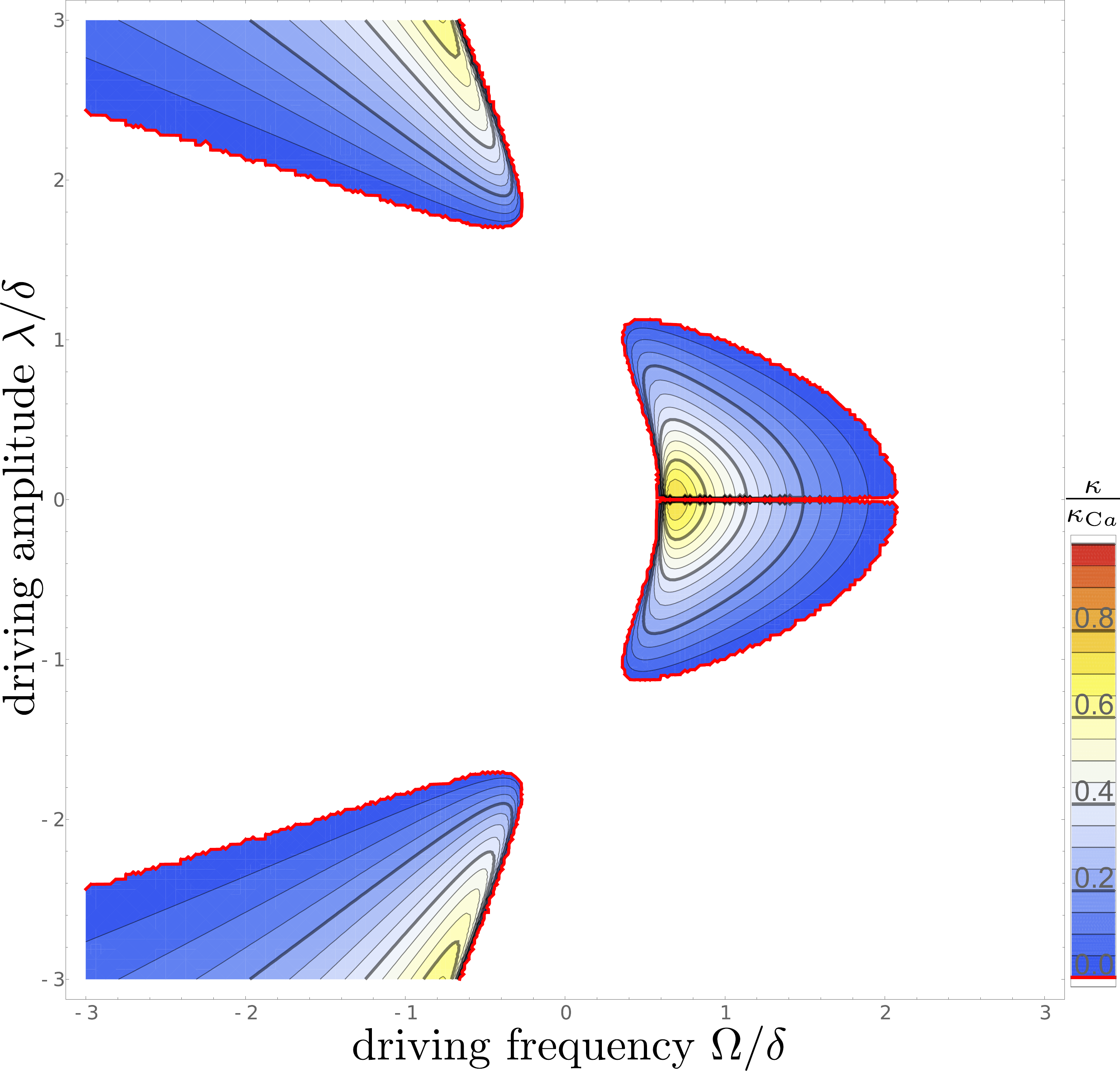}
    \caption{Analogous to Fig.~\ref{FIG:current_lb_1qutrit}, but plotting the renormalized coefficient of performance for cooling~\eqref{EQ:cop} instead, with contours in \rev{steps of $0.05$}. The maximum coefficient of performance is bound by its Carnot limit. Regions of maximum coefficient of performance do not coincide with regions of maximal current. In contrast, the weak-driving master equation would yield a constant value $1/2$ throughout the plotting region.}
    \label{FIG:cop_lb_1qutrit}
\end{figure}

From the previous subsection we also see that the dissipator of the Floquet Lindblad master equation $\dot{\f{\rho}} = {\cal L}\f{\rho}$ additively decomposes into hot and cold reservoir contributions, with each dissipator leading to local thermalization at a shifted steady state $\bar\rho_\nu$
\begin{align}
    {\cal L} &= {\cal L}_c + {\cal L}_h\,,\qquad
    {\cal L}_\nu \bar\rho_\nu = 0\,,\qquad \bar\rho_\nu = \frac{e^{-\beta_\nu H_S^\nu}}{Z_\nu}\,,\nn
    H_S^h &\equiv (\epsilon_++\frac{\Omega}{2})\sum_i (\ket{+}\bra{+})_i\nn
    &\qquad+ (\epsilon_-+\frac{\Omega}{2})\sum_i (\ket{-}\bra{-})_i\,,\nn
    H_S^c &\equiv (\epsilon_+-\frac{\Omega}{2})\sum_i (\ket{+}\bra{+})_i\nn
    &\qquad+ (\epsilon_--\frac{\Omega}{2})\sum_i (\ket{-}\bra{-})_i\,,
\end{align}
whereas the ground state energies of the bare Hamiltonian $\epsilon_0=0$ remain untouched.
This means that the $H_S^\nu$ are not just shifted versions of the collective Floquet Hamiltonian, and the reservoir-specific steady states $\bar\rho_\nu$ are not just Floquet-Gibbs states~\cite{shirai2015a,shirai2016a}.
Comparing with the counting field-based energy current~\eqref{EQ:current_lindblad2}, one could therefore alternatively define the heat currents by using the effective reservoir-specific Hamiltonians
\begin{align}
I_E^\nu = \trace{H_S^\nu ({\cal L}^\nu \f{\rho)}}\,,    
\end{align}
and the power at steady state by invoking the first law $\bar P = - \bar I_E^c - \bar I_E^h$.
With this definition, we can write for the entropy change of the system
\begin{align}
    \dot S &= -\trace{(\sum_\nu {\cal L}_\nu \rho) \ln \rho}\nn
    &= -\trace{(\sum_\nu {\cal L}_\nu \rho) [\ln \rho-\ln\bar\rho_\nu]}+ \sum_\nu \beta_\nu I_E^\nu\nn
    &\equiv \dot{\sigma}_\ii + \sum_\nu \beta_\nu I_E^\nu\,,
\end{align}
where by Spohn's inequality~\cite{spohn1978b} we can infer that the irreversible entropy production is positive $\dot{\sigma}_\ii\ge 0$.
Solving for it, we thus find the usual form for the entropy production rate
\begin{align}
    \dot{\sigma}_\ii = \dot S - \sum_\nu \beta_\nu I_E^\nu \stackrel{t\to\infty}{\to} -\sum_\nu \beta_\nu \bar I_E^\nu \ge 0\,,
\end{align}
which at steady-state bounds the currents of the Floquet-Lindblad master equation.
Eventually, this also bounds the coefficient of performance of our device by its Carnot value, analogous to the discussion in Sec.~\ref{SEC:thermodynamics_adiabatic}.
For the parameters in Fig.~\ref{FIG:current_lb_1qutrit}, we demonstrate this by showing the renormalized coefficient of performance~\eqref{EQ:cop} in Fig.~\ref{FIG:cop_lb_1qutrit}, which is always below one.
In contrast, for the weak-driving master equation, the coefficient of performance is constant $\delta/(\Delta-\delta)$ as discussed in Sec.~\ref{SEC:thermodynamics_adiabatic}.
As usual, regions of maximum cooling current do not coincide with regions of best coefficient of performance, such that finding the optimal operational regime is a tradeoff between these properties.

However, if in addition the standard thermodynamic uncertainty relation~\eqref{EQ:tur} is also respected, the coefficient of performance can be bound tighter
\begin{align}
    \kappa &= \frac{\bar I_E^c}{-\bar I_E^c-\bar I_E^h} = \frac{\beta_h \bar I_E^c}{\dot{\bar{\sigma}}_\ii + (\beta_c-\beta_h) \bar I_E^c}\\
    &\le  \frac{\beta_h \bar I_E^c}{2 \frac{(\bar I_E^c)^2}{\bar S_E^c} + (\beta_c-\beta_h)\bar I_E^c} = \kappa_{\rm Ca} \frac{1}{1+2 \frac{\bar I_E^c}{\bar S_E^c (\beta_c-\beta_h)}}\nonumber
\end{align}
below the Carnot coefficient of performance.

\subsection{Case of a single qutrit}

For a single qutrit, Eq.~\eqref{EQ:pauli_N} reduces into
\begin{align}\label{EQ:pauli_1}
    \dot P_0 &= \left[\cos^2(\alpha) \gamma_c\left(\epsilon_--\frac{\Omega}{2}\right)+\sin^2(\alpha)\gamma_h\left(\epsilon_-+\frac{\Omega}{2}\right)\right] P_-\nn
    &\quad+\left[\sin^2(\alpha)\gamma_c\left(\epsilon_+-\frac{\Omega}{2}\right)+\cos^2(\alpha)\gamma_h\left(\epsilon_++\frac{\Omega}{2}\right)\right] P_+\nn
    &\qquad- [\ldots] P_0\,,\nn
    \dot P_- &= \left[\cos^2(\alpha) \gamma_c\left(-\epsilon_-+\frac{\Omega}{2}\right)+\sin^2(\alpha) \gamma_h\left(-\epsilon_--\frac{\Omega}{2}\right)\right] P_0\nn
    &\quad- [\ldots] P_-\,,\nn
    \dot P_+ &= \left[\sin^2(\alpha) \gamma_c\left(-\epsilon_++\frac{\Omega}{2}\right)+\cos^2(\alpha) \gamma_h\left(-\epsilon_+-\frac{\Omega}{2}\right)\right] P_0\nn
    &\quad- [\ldots] P_+\,,
\end{align}
\rev{where the diagonal $[\ldots]$-terms are fixed by the probability conservation -- rate matrices must have vanishing column sum.}
This rate equation system hosts two cycles shown in Fig.~\ref{FIG:coolingconfig} \rev{with effective transition energies} 
that do not comply with the Floquet energies, such that under appropriate conditions one may reach cooling of the cold reservoir.

\section{Floquet-Redfield master equation}\label{APP:redfield}

\rev{In this appendix, we consider a non-Lindblad master equation for our system obtained from a perturbative treatment of the system-reservoir coupling strength $B^\nu$. We show how the asymptotically periodic long-term solution and resulting period-averaged current can be obtained and in particular discuss the difference between the Floquet-Redfield and the Floquet-Lindblad currents, making formulas explicit for the case $N=1$.}

\subsection{Derivation} 

Starting from~\eqref{EQ:redfield2}, we avoid the secular approximation and directly switch back to the Schr\"odinger picture \rev{by introducing the $t$- and $\tau$-dependent operator}
$\tilde J^\nu_\rev{t\tau} \equiv U_S(t) U_S^\dagger(t-\tau) J^\nu U_S(t-\tau) U_S^\dagger(t)$.
\rev{This yields}
\begin{align}\label{EQ:redfield3}
    \dot{\rho} &= -\ii [H_S(t), \rho]-\Big\{\sum_\nu \int_0^\infty d\tau C_\nu(+\tau) \left[J^\nu, \tilde J^\nu_\rev{t\tau} \rho\right]\nn
    &\qquad+{\rm h.c.}\Big\}\,,
\end{align}
\rev{and by writing $U_S(t)= e^{-\ii H_S^0 t} \tilde U_S(t)$ with $\dot{\tilde{U}}_S(t)=-\ii e^{+\ii H_S^0 t} V(t) e^{-\ii H_S^0 t} \tilde U_S(t)$ one finds with $\tilde J^\nu_\rev{t\tau} = e^{-\ii H_S^0 \tau} J^\nu e^{+\ii H_S^0 \tau} + \ord\{\lambda\}$ that the above equation would for small driving strengths $\lambda$ also yield a Redfield version of the weak-driving master equation discussed in App.~\ref{APP:weakdriving}.
However, the collective time evolution operator can be computed for all driving strengths $\lambda$, which leads to 
} 
\begin{align}
    \tilde J^c_\rev{t\tau} &= \sum_{a\in\pm} e^{+\ii(\epsilon_a-\frac{\Omega}{2})\tau} \left[\abs{\braket{1}{a}}^2 J_c^- + \braket{1}{a}\braket{a}{2} e^{+\ii\Omega t} J_h^-\right]\nn
    &+\sum_{a\in\pm} e^{-\ii(\epsilon_a-\frac{\Omega}{2})\tau} \left[\abs{\braket{1}{a}}^2 J_c^+ + \braket{2}{a}\braket{a}{1} e^{-\ii\Omega t} J_h^+\right]\,,\nn
    \tilde J^h_\rev{t\tau} &= \sum_{a\in\pm} e^{+\ii(\epsilon_a+\frac{\Omega}{2})\tau} \left[\abs{\braket{2}{a}}^2 J_h^- + \braket{2}{a}\braket{a}{1} e^{-\ii\Omega t} J_c^-\right]\nn
    &+\sum_{a\in\pm} e^{-\ii(\epsilon_a+\frac{\Omega}{2})\tau} \left[\abs{\braket{2}{a}}^2 J_h^+ + \braket{1}{a}\braket{a}{2} e^{+\ii\Omega t} J_c^+\right]\,.
\end{align}
Then, we can insert the Fourier transform of the correlation functions $C_\nu(\tau)=\frac{1}{2\pi} \int \gamma_\nu(\omega) e^{-\ii\omega\tau} d\omega$, perform the $\int d\tau$ integrations and invoke the Sokhotskij-Plemelj theorem~\eqref{EQ:spt} while neglecting the Lamb-shift $\frac{1}{2\pi} \int_0^\infty e^{+\ii\omega\tau} d\tau \approx \frac{1}{2} \delta(\omega) $, to arrive at Eq.~\eqref{EQ:redfield} in the main text.

\begin{figure}
    \includegraphics[width=0.46\textwidth]{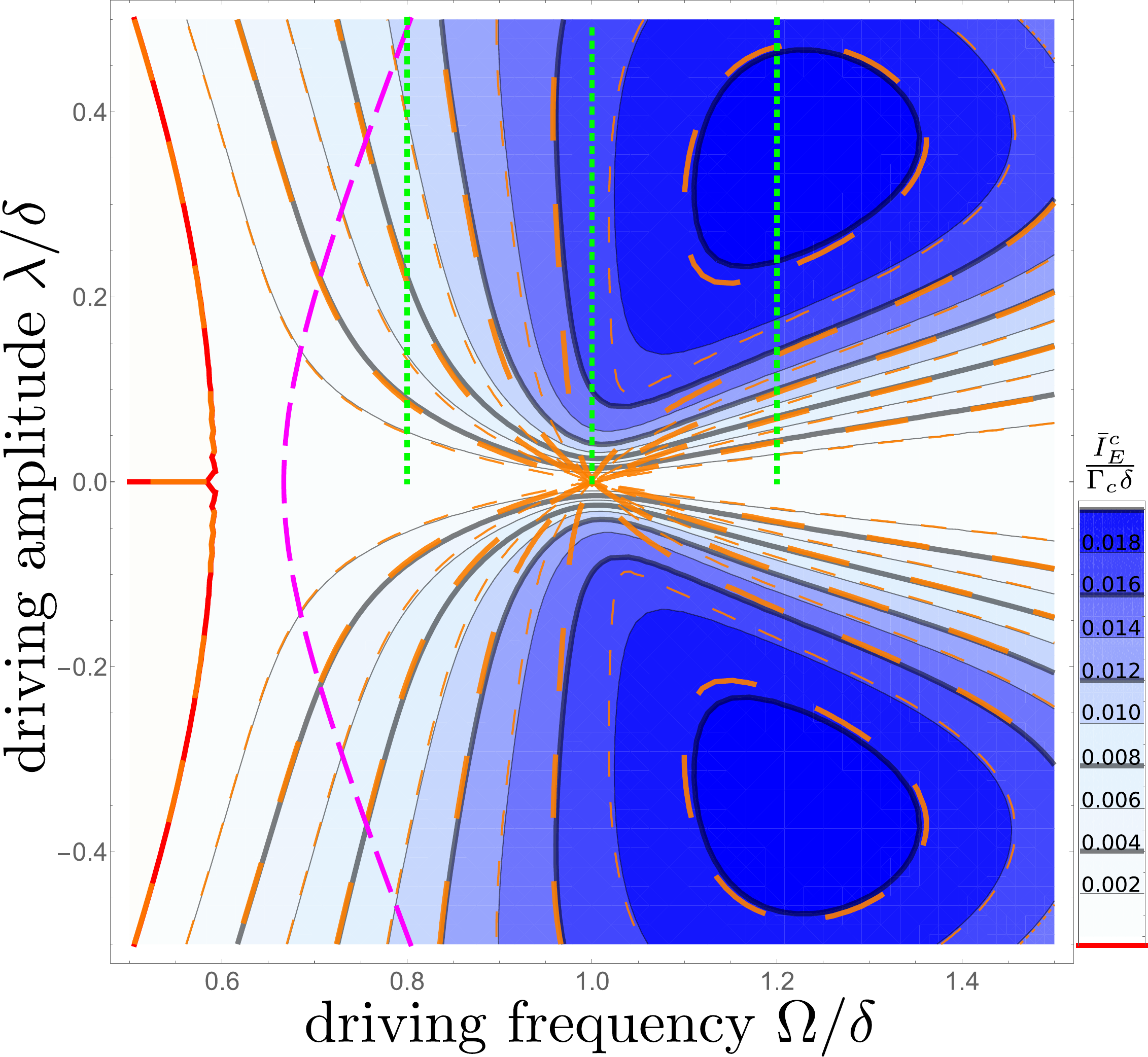}
    \caption{Analogous to Fig.~\ref{FIG:current_lb_1qutrit} (in colour coding \rev{and} parameters), but plotting the period-averaged Redfield current~\eqref{EQ:pa_current_rf_cold} around the region where the Floquet-Lindblad description fails $\Omega/\delta\in[0.5,1.5]$ and $\lambda/\delta\in[-0.5,+0.5]$. Dashed orange contours correspond to the Floquet-Lindblad contours and differ from the Floquet-Redfield contours (black) only near resonance $\Omega\approx \Delta-\delta$ and $\lambda\approx 0$. The Redfield current vanishes identically along the line $\lambda=0$ (\rev{the associated horizontal red contour line has been removed} to improve visibility).}
    \label{FIG:current_rf_1qutrit}
\end{figure}

In the Schr\"odinger picture, we thereby obtain a time-dependent generator, and its Fourier components are
\begin{align}\label{EQ:sidebandsexp}
    {\cal L}_0 \rho &= -\ii [H_S^0, \rho]\nn
    &\qquad-\sum_a \frac{\gamma_c(-\epsilon_a+\frac{\Omega}{2})\abs{\braket{a}{1}}^2}{2} \left\{\left[J^c, J_c^+ \rho\right] + {\rm h.c.}\right\}\nn
    &\qquad-\sum_a \frac{\gamma_c(+\epsilon_a-\frac{\Omega}{2})\abs{\braket{a}{1}}^2}{2} \left\{\left[J^c, J_c^- \rho\right] + {\rm h.c.}\right\}\nn
    &\qquad-\sum_a \frac{\gamma_h(-\epsilon_a-\frac{\Omega}{2})\abs{\braket{a}{2}}^2}{2} \left\{\left[J^h, J_h^+ \rho\right] + {\rm h.c.}\right\}\nn
    &\qquad-\sum_a \frac{\gamma_h(+\epsilon_a+\frac{\Omega}{2})\abs{\braket{a}{2}}^2}{2} \left\{\left[J^h, J_h^- \rho\right] + {\rm h.c.}\right\}\,,\nn
    {\cal L}_+ \rho &= -\ii [\lambda J_w^-, \rho]\nn
    &\qquad -\sum_a \frac{\gamma_c(-\epsilon_a+\Omega/2)\braket{1}{a}\braket{a}{2}}{2} \left[\rho J_h^-, J^c\right]\nn
    &\qquad -\sum_a \frac{\gamma_c(+\epsilon_a-\Omega/2)\braket{1}{a}\braket{a}{2}}{2} \left[J^c, J_h^- \rho\right]\nn
    &\qquad -\sum_a \frac{\gamma_h(-\epsilon_a-\Omega/2)\braket{1}{a}\braket{a}{2}}{2} \left[J^h, J_c^+ \rho\right]\nn
    &\qquad -\sum_a \frac{\gamma_h(+\epsilon_a+\Omega/2)\braket{1}{a}\braket{a}{2}}{2} \left[\rho J_c^+, J^h\right]\,,\nn
    {\cal L}_- \rho &= -\ii [\lambda^* J_w^+, \rho]\\
    &\qquad -\sum_a \frac{\gamma_c(-\epsilon_a+\Omega/2)\braket{2}{a}\braket{a}{1}}{2} \left[J^c, J_h^+ \rho\right]\nn
    &\qquad -\sum_a \frac{\gamma_c(+\epsilon_a-\Omega/2)\braket{2}{a}\braket{a}{1}}{2} \left[\rho J_h^+, J^c\right]\nn
    &\qquad -\sum_a \frac{\gamma_h(-\epsilon_a-\Omega/2)\braket{2}{a}\braket{a}{1}}{2} \left[\rho J_c^-, J^h\right]\nn
    &\qquad -\sum_a \frac{\gamma_h(+\epsilon_a+\Omega/2)\braket{2}{a}\braket{a}{1}}{2} \left[J^h, J_c^- \rho\right]\,.\nonumber
\end{align}
Using a matrix representation of the superoperators, the asymptotic periodic state is then found as the (trace-normalized) nullspace of the infinite-dimensional matrix
\begin{align}
\left(\begin{array}{c|ccc|c}
    \ddots & \ddots & & & \\
    \hline
    {\cal L}_+ & ({\cal L}_0 + \ii \Omega\f{1}) & {\cal L}_- & &\\
    & {\cal L}_+ & ({\cal L}_0) & {\cal L}_- &\\
    & & {\cal L}_+ & ({\cal L}_0 - \ii \Omega\f{1}) & {\cal L}_-\\
    \hline
    & & & \ddots& \ddots
\end{array}\right)\,,
\end{align}
where in general a suitable cutoff (in the example above indicated by the lines as $n_{\rm cut}=\pm1$) has to be chosen such that convergence is ensured.
Based on the replacements~\eqref{EQ:replacement3}, the period-averaged current~\eqref{EQ:pa_current_rf} then becomes
\begin{align}\label{EQ:pa_current_rf_cold}
    \bar I_E^c &= \sum_a \frac{\epsilon_a-\Omega/2}{2} \abs{\braket{a}{1}}^2\times\nn
    &\qquad\times\Big[\gamma_c(-\epsilon_a+\Omega/2) \trace{(J^c J_c^+ + J_c^- J^c)\bar\rho^{(0)}}\nn
    &\qquad\qquad- \gamma_c(+\epsilon_a-\Omega/2) \trace{(J^c J_c^- + J_c^+ J^c)\bar\rho^{(0)}}\Big]\nn
    &\qquad+\sum_a \frac{\epsilon_a-\Omega/2}{2} \braket{1}{a}\braket{a}{2}\times\nn
    &\qquad\times\Big[\gamma_c(-\epsilon_a+\Omega/2) \trace{J_h^- J^c \bar\rho^{(-1)}}\nn
    &\qquad\qquad- \gamma_c(\epsilon_a-\Omega/2) \trace{J^c J_h^- \bar\rho^{(-1)}}\Big]\nn
    &\qquad+\sum_a \frac{\epsilon_a-\Omega/2}{2} \braket{2}{a}\braket{a}{1}\times\nn
    &\qquad\times\Big[\gamma_c(-\epsilon_a+\Omega/2) \trace{J^c J_h^+ \bar\rho^{(+1)}}\nn
    &\qquad\qquad- \gamma_c(\epsilon_a-\Omega/2) \trace{J_h^+ J^c \bar\rho^{(+1)}}\Big]\,.
\end{align}
When we plot this in analogy to Fig.~\ref{FIG:current_lb_1qutrit} in the main text, we obtain a quite similar result, see Fig.~\ref{FIG:current_rf_1qutrit}.
The difference however is that the Redfield equation also applies to the case where the secular approximations fail, such that we indeed find a vanishing current along the line $\lambda=0$, in contrast to Fig.~\ref{FIG:current_lb_1qutrit} in the main text.

Analogous considerations for the hot reservoir lead to
\begin{align}
    \bar I_E^h &= \sum_a \frac{\epsilon_a+\Omega/2}{2} \abs{\braket{a}{2}}^2\times\nn
    &\qquad\times\Big[\gamma_h(-\epsilon_a-\Omega/2) \trace{(J^h J_h^+ + J_h^- J^h)\bar\rho^{(0)}}\nn
    &\qquad\qquad- \gamma_h(+\epsilon_a+\Omega/2) \trace{(J^h J_h^- + J_h^+ J^h)\bar\rho^{(0)}}\Big]\nn
    &\qquad+\sum_a \frac{\epsilon_a+\Omega/2}{2} \braket{1}{a}\braket{a}{2}\times\nn
    &\qquad\times\Big[\gamma_h(-\epsilon_a-\Omega/2) \trace{J^h J_c^+ \bar\rho^{(-1)}}\nn
    &\qquad\qquad- \gamma_h(\epsilon_a+\Omega/2) \trace{J_c^+ J^h \bar\rho^{(-1)}}\Big]\nn
    &\qquad+\sum_a \frac{\epsilon_a+\Omega/2}{2} \braket{2}{a}\braket{a}{1}\times\nn
    &\qquad\times\Big[\gamma_h(-\epsilon_a-\Omega/2) \trace{J_c^- J^h \bar\rho^{(+1)}}\nn
    &\qquad\qquad- \gamma_h(\epsilon_a+\Omega/2) \trace{J^h J_c^-\bar\rho^{(+1)}}\Big]\,.
\end{align}

\subsection{Case of a single qutrit}

For a single qutrit, we can obtain from Eq.~\eqref{EQ:redfield} in the main text coupled equations for the matrix elements of the system density matrix.
From their structure (in the Schr\"odinger picture and the original basis) we already see that the the trace is conserved $\dot\rho_{11}+\dot\rho_{22}+\dot\rho_{33}=0$ and that the density matrix remains hermitian throughout
\begin{align}
    \dot \rho_{00} &= -(L_{0\to 1}+L_{0\to 2}) \rho_{00} + L_{1\to 0} \rho_{11} + L_{2\to 0} \rho_{22}\nn
    &\qquad + (L_{\downarrow c}+L_{\downarrow h}) e^{-\ii\Omega t} \rho_{12} + (L_{\downarrow c}^*+L_{\downarrow h}^*) e^{+\ii\Omega t} \rho_{21}\,,\nn
    \dot \rho_{11} &= L_{0\to 1} \rho_{00} - L_{1\to 0} \rho_{11}\nn
    &\qquad -(L_{\downarrow c} - \ii\lambda^*) e^{-\ii\Omega t} \rho_{12} - (L_{\downarrow c}^*+\ii\lambda) e^{+\ii\Omega t} \rho_{21}\,,\nn
    \dot \rho_{22} &= L_{0\to 2} \rho_{00} - L_{2\to 0} \rho_{22}\nn
    &\qquad -(L_{\downarrow h} + \ii\lambda^*) e^{-\ii\Omega t} \rho_{12} - (L_{\downarrow h}^*-\ii\lambda) e^{+\ii\Omega t} \rho_{21}\,,\nn
    \dot \rho_{12} &= (L_{\uparrow c}^* + L_{\uparrow h}^*) e^{+\ii\Omega t} \rho_{00} - (L_{\downarrow h}^* - \ii \lambda) e^{+\ii\Omega t} \rho_{11}\nn
    &\qquad- (L_{\downarrow c}^* + \ii \lambda) e^{+\ii\Omega t} \rho_{22}\nn
    &\qquad- (L_{1\to 0}/2 + L_{2\to 0}/2- \ii \Delta + \ii \delta) \rho_{12}\,,\nn
    \dot \rho_{21} &= (L_{\uparrow c} + L_{\uparrow h}) e^{-\ii\Omega t} \rho_{00} - (L_{\downarrow h} + \ii \lambda^*) e^{-\ii\Omega t} \rho_{11}\nn
    &\qquad- (L_{\downarrow c}- \ii \lambda^*) e^{-\ii\Omega t} \rho_{22}\nn
    &\qquad- (L_{1\to 0}/2 + L_{2\to 0}/2 + \ii \Delta - \ii \delta) \rho_{21}\,,
\end{align}
where
\begin{align}
    L_{1\to 0} &= \sum_a \abs{\braket{a}{1}}^2  \gamma_c(\epsilon_a-\Omega/2)\,,\nn
    L_{0\to 1} &= \sum_a \abs{\braket{a}{1}}^2  \gamma_c(-\epsilon_a+\Omega/2)\,,\nn
    L_{2\to 0} &= \sum_a \abs{\braket{a}{2}}^2  \gamma_h(\epsilon_a+\Omega/2)\,,\nn    
    L_{0\to 2} &= \sum_a \abs{\braket{a}{2}}^2  \gamma_h(-\epsilon_a-\Omega/2)\,,\nn
    L_{\downarrow c} &= \sum_a \braket{2}{a}\braket{a}{1}/2 \gamma_c(\epsilon_a-\Omega/2)\,,\nn
    L_{\uparrow c} &= \sum_a \braket{2}{a}\braket{a}{1}/2 \gamma_c(-\epsilon_a+\Omega/2)\,,\nn
    L_{\downarrow h} &= \sum_a \braket{2}{a}\braket{a}{1}/2 \gamma_h(\epsilon_a+\Omega/2)\,,\nn
    L_{\uparrow h} &= \sum_a \braket{2}{a}\braket{a}{1}/2 \gamma_h(-\epsilon_a-\Omega/2)\,.
\end{align}
The Hamiltonian part of this equation, defined by $\Delta$, $\delta$, and $\lambda$, naturally already agrees with the weak-driving master equation~\eqref{EQ:lindblad_adiabatic}.
When $\lambda\to 0$, one can show that the other parts also fall back to that under the secular approximation: For small $\lambda$, the eigenstates of the Floquet Hamiltonian approach the eigenstates of the undriven system Hamiltonian, and we get 
$L_{1\to 0} \to \gamma_c(\delta)$, $L_{0\to 1} \to \gamma_c(-\delta)$,  
$L_{2\to 0} \to \gamma_h(\Delta)$, $L_{0\to 2} \to \gamma_h(-\Delta)$, whereas the other matrix elements vanish $L_{\updownarrow \nu} \to 0$.
However, a Taylor expansion of the dissipative terms to first order in $\lambda$ shows that the above single-qutrit Redfield equation and the weak-driving master equation~\eqref{EQ:lindblad_adiabatic} are not fully equivalent. 
They become equivalent when we drop the oscillatory terms in the dissipators (equivalent to the secular approximation).

For a single qutrit we obtain for the current leaving the cold reservoir the expression 
\begin{align}
\bar I_E^c  &=-\sum_{a\in\pm} (\epsilon_a-\Omega/2) \gamma_c(\epsilon_a-\Omega/2) \abs{\braket{a}{1}}^2 \bar\rho_{11}^{(0)}\nn
    &\qquad-\sum_{a\in\pm} (-\epsilon_a+\Omega/2) \gamma_c(-\epsilon_a+\Omega/2) \abs{\braket{a}{1}}^2 \bar\rho_{00}^{(0)}\nn
    &\qquad-\sum_{a\in\pm} (\epsilon_a-\Omega/2) \frac{\gamma_c(\epsilon_a-\Omega/2)}{2}\Big[\braket{1}{a} \braket{a}{2} \bar\rho_{21}^{(-1)}\nn
    &\qquad\qquad+ \braket{2}{a}\braket{a}{1} \bar\rho_{12}^{(+1)}\Big]\,,
\end{align}
which is defined positive in the regime of cooling functionality.

\section{Full Counting Statistics}\label{APP:fcs}

We are interested in the flow of energy out of or into the reservoirs and therefore we sketch 
the microscopic derivation of the counting field formalism from Ref.~\cite{esposito2009a} \rev{specific to the statistics of energy exchanges here.}
The method can be used to extract also higher moments like noise, but we are most interested in the current here. 
Furthermore, we consider just one reservoir here, but the method can be extended to multiple ones in a straightforward way.
Also generalizations to other observables than the bath energy like particle number or spin are possible, the only requirement being the assumption that the bath observable commutes with the bath Hamiltonian. 

The scheme employs a two-point measurement scheme, where at time $t=0$ we measure the energy of the reservoir, leading to outcome 
$E_\ell$ after which -- with the spectral representation $H_B = \sum_\ell E_\ell \ket{\ell}\bra{\ell}$ -- the reservoir density matrix is projected onto
\begin{align}
    \bar\rho_B \to  \frac{\ket{\ell}\bra{\ell}\bar\rho_B\ket{\ell}\bra{\ell}}{P_\ell} \equiv \frac{\bar\rho_B^{(\ell)}}{P_\ell}\,,
\end{align}
where $P_\ell = \trace{\ket{\ell}\bra{\ell} \bar\rho_B} = \bra{\ell}\bar\rho_B\ket{\ell}$ is the probability of obtaining this initial outcome.
In the subsequent evolution, the energy content of the reservoir changes with respect to this initial value, and an exact moment-generating function (MGF) for the change of the reservoir energy, averaged over all initial outcomes $\ell$, can be given as
\begin{align}
    M(\chi,t) = \sum_\ell \trace{e^{\ii\chi (H_B-E_\ell)} \f{U}(t) \rho_S^0 \bar\rho_B^{(\ell)} \f{U^\dagger}(t)}\,,
\end{align}
where we have used an interaction picture (denoted by bold symbols) with respect to $H_B$, such that the observable does not pick up any time-dependence, and where $\f{U}$ denotes the full time evolution operator.
By performing derivatives $(-\ii\partial_\chi)^k$ and setting $\chi\to 0$ afterwards one thus obtains moments of the energy changes in the reservoir.

The MGF can be rewritten as
\begin{align}\label{EQ:mgf_full}
    M(\chi,t) = \trace{\f{U_{+\chi/2}}(t) \rho_S^0 \otimes \bar\rho_B \f{U_{-\chi/2}^\dagger}(t)}\,,
\end{align}
where we used $\sum_\ell \bar\rho_B^{(\ell)} = \bar\rho_B$ and introduced a generalized time evolution operator
$\f{U_{+\chi/2}}(t) = e^{+\ii H_B \chi/2} \f{U}(t) e^{-\ii H_B \chi/2}$, which is just the solution to
$\f{\dot{U}_{+\chi/2}}(t) = -\ii \f{H_I^{\chi/2}}(t)  \f{U_{+\chi/2}}(t)$, with the tilted interaction Hamiltonian (we assume for brevity a single-operator decomposition $H_I = S \otimes B$ like used in the main text)
\begin{align}
    \f{H_I^{\chi/2}}(t) &= e^{+\ii H_B \chi/2} \f{H_I}(t) e^{-\ii H_B \chi/2}\nn
    &= \f{S}(t) \otimes e^{+\ii H_B \chi/2} \f{B}(t) e^{-\ii H_B \chi/2}\,.
\end{align}
Accordingly, the time-dependence of the reservoir coupling operators is just shifted, and by applying a perturbative scheme on top (like e.g. the microscopic derivations discussed before), one can obtain a 
perturbative approximation to the tilted system density matrix 
$\f{\rho}(\chi,t) = \traceB{\f{U_{+\chi/2}}(t) \rho_S^0 \otimes \bar\rho_B \f{U_{-\chi/2}^\dagger}(t)}$
and the derived MGF $M(\chi,t)=\traceS{\f{\rho}(\chi,t)}$.
The tilted density matrix $\rho(\chi,t)$ falls back to the standard system density matrix for $\chi=0$ but in addition now allows to extract (approximations to) the desired moments of the reservoir energy changes via performing suitable derivatives
$\expval{\Delta E_B^k} = (-\ii \partial_\chi)^k \trace{\rho(\chi,t)}|_{\chi=0}$.
To second order in the system-bath interaction, the modification of the bath coupling operators will only affect terms where the system density matrix is sandwiched by two system coupling operators: There, one of them is transformed with $+\chi$, whereas the other is transformed with $-\chi$, which gives rise to shifted correlation functions
\begin{align}
    \traceB{\f{B}(t+\chi/2) \bar\rho_B \f{B}(t'-\chi/2)} &= C(t'-t-\chi)\,,\nn
    \traceB{\f{B}(t'+\chi/2) \bar\rho_B \f{B}(t-\chi/2)} &= C(t-t'-\chi)\,,
\end{align}
whose Fourier transforms are therefore multiplied by a phase
$\gamma(\omega) \to \gamma(\omega) e^{+\ii\omega\chi}$.
In the terms where $\rho$ is not sandwiched, the bath coupling operators are transformed with the same $\chi$, such that the counting field cancels and one obtains the standard correlation function.
Following the very same procedures as in the microscopic derivations outlined before, this eventually leads to the suggested replacements in Eqns.~\eqref{EQ:replacement1},~\eqref{EQ:replacement2}, and~\eqref{EQ:replacement3} of the main text.

The energy current entering the reservoir is nothing but the time derivative of the first moment
\begin{align}
    I_{E,B} &= \frac{d}{dt} (-\ii \partial_\chi) \trace{\rho(\chi,t)}|_{\chi=0}\nn
    &= (-\ii \partial_\chi) \trace{{\cal L}(\chi,t) \rho(\chi,t)}|_{\chi=0}\nn
    &= -\ii \trace{[\partial_\chi {\cal L}(\chi,t)|_{\chi=0}] \rho(0,t)}\,,
\end{align}
where we used that ${\cal L}(0,t)$ is trace-conserving.
In the main text we merely changed the sign of the above current formula to align with the convention that the cooling current should count positive when decreasing the reservoir energy.
Also simplified formulas for higher moments can be obtained for driven systems~\cite{benito2016b,restrepo2019a}.

\section{Collective bases}\label{APP:collective_bases}

To acknowledge the exact permutational symmetry of our system \rev{for many qutrits, we also work in a permutationally symmetric basis.
The matrix elements in this basis can be expressed with non-standard Clebsch-Gordan coefficients that can be obtained from bosonization techniques of the Holstein-Primakoff type~\cite{klein1991a,providencia2006a}.
We mainly state the action of the ladder operators here but more information can be found in the appendices of Ref.~\cite{kolisnyk2023a}.
}

We define the normalized permutationally symmetric states of $N$ qutrits with $M$ large and $m$ small excitations as 
\begin{align}
    \ket{M;m} \propto (J_h^+)^M (J_c^+)^m \ket{0\ldots 0}\,,
\end{align}
where $\ket{0;0} = \ket{0\ldots 0}$ is the state without any excitations, $\ket{N;0}=\ket{2\ldots 2}$ the state with all qutrits in their most excited state and $\ket{0;N}$ is the state with all qutrits in their first excited state, see also Fig.~\ref{FIG:threels4_naive}.
In the fully symmetric subspace, these two excitations can be directly mapped to two bosonic modes~\cite{klein1991a}, from which we get the relations
\begin{align}\label{EQ:ladder_collective}
    J_h^+ \ket{M;m} &= \sqrt{(N-M-m)(M+1)} \ket{M+1;m}\,,\nn
    J_h^- \ket{M;m} &= \sqrt{(N-M-m+1)M} \ket{M-1;m}\,,\nn
    J_c^+ \ket{M;m} &= \sqrt{(N-M-m)(m+1)} \ket{M;m+1}\,,\nn
    J_c^- \ket{M;m} &= \sqrt{(N-M-m+1)m} \ket{M;m-1}\,,\nn
    J_w^+ \ket{M;m} &= \sqrt{(M+1)m} \ket{M+1;m-1}\,,\nn
    J_w^- \ket{M;m} &= \sqrt{M(m+1)} \ket{M-1;m+1}\,.
\end{align}

As the eigenstates~\eqref{EQ:floquet_states} of the single-particle Floquet Hamiltonian~\eqref{EQ:hamfloquet} are just rotations of the original basis states, we may likewise introduce the collectively rotated ladder operators
\begin{align}\label{EQ:rottransform}
    S_-^- &= \sum_i (\ket{0}\bra{-})_i = \cos(\alpha) J_c^- - \sin(\alpha) J_h^-\,,\nn
    S_+^- &= \sum_i (\ket{0}\bra{+})_i = \sin(\alpha) J_c^- + \cos(\alpha) J_h^-\,,
\end{align}
which can be used to construct the collective permutationally symmetric Floquet eigenstates
(which we denote by a comma instead) from the state with no excitations
\begin{align}
    \ket{M,m} \propto (S_+^+)^M (S_-^+)^m \ket{0\ldots 0}\,,
\end{align}
see also Fig.~\ref{FIG:threels4_driven}.
By construction, these obey the eigenvalue equation
\begin{align}
    H_F^{\rm coll} \ket{M,m} &= (M\epsilon_+ + m \epsilon_-) \ket{M,m}
\end{align}
with the collective Floquet Hamiltonian $H_F^{\rm coll} = \sum_i (H_F)_i = \Delta \sum_i (\ket{2}\bra{2})_i + \delta \sum_i (\ket{1}\bra{1})_i$ and Floquet energies~\eqref{EQ:enfloquet}.

With this, also the matrix elements in the collective Floquet eigenbasis can be conveniently evaluated by using the conventional ladder operator properties~\eqref{EQ:ladder_collective}, e.g.
\begin{align}\label{EQ:collective_matrix_element}
    \bra{M,m-1} S_-^- \ket{M,m} &= \bra{M;m-1} J_c^- \ket{M;m}\nn
    &= \sqrt{(N-M-m+1)m}\,,
\end{align}
which is used to calculate the matrix elements occurring in the Floquet-Pauli rate equation~\eqref{EQ:pauli_N}.

\bibliographystyle{unsrt}
\bibliography{references}

\end{document}